\documentclass[reprint,amsmath,amssymb,aps,pra,longbibliography,showkeys,floatfix]{revtex4-2}

\usepackage{amsmath,amssymb}
\usepackage{graphicx}
\usepackage{booktabs}
\usepackage{microtype}
\usepackage{hyperref}
\hypersetup{hidelinks}
\usepackage{xcolor}

\newcommand{\DEadvB}{A_{\mathrm{DE,BFGS}}}
\newcommand{\DEadvL}{A_{\mathrm{DE,local}}}

\newcommand{\maintablestyle}{%
  \small
  \renewcommand{\arraystretch}{1.10}%
  \setlength{\tabcolsep}{4.5pt}%
}
\newcommand{\mhead}[1]{\shortstack[c]{#1}}
\newcommand{\paperfig}[2]{%
  \IfFileExists{figures/#1}{\includegraphics[width=#2]{figures/#1}}%
  {\fbox{\parbox[c][4.5cm][c]{#2}{\centering Upload \texttt{\detokenize{#1}} to \texttt{figures/}}}}%
}

\begin{document}

\title{When is global evolutionary search useful for variational quantum algorithms? A landscape-first study}

\author{Vojt\v{e}ch Nov\'{a}k}
\email{vojtech.novak.st1@vsb.cz}
\affiliation{Department of Computer Science, Faculty of Electrical Engineering and Computer Science, VSB - Technical University of Ostrava, Ostrava, Czech Republic}
\affiliation{IT4Innovations National Supercomputing Center, VSB - Technical University of Ostrava, 708 00 Ostrava, Czech Republic}
\affiliation{Department of Informatics and Statistics, Marine Research Institute, Klaipeda University, Lithuania}

\author{Ivan Zelinka}
\affiliation{Department of Computer Science, Faculty of Electrical Engineering and Computer Science, VSB - Technical University of Ostrava, Ostrava, Czech Republic}
\affiliation{IT4Innovations National Supercomputing Center, VSB - Technical University of Ostrava, 708 00 Ostrava, Czech Republic}
\affiliation{Department of Informatics and Statistics, Marine Research Institute, Klaipeda University, Lithuania}

\begin{abstract}
Variational quantum algorithms turn quantum-state preparation into a classical nonconvex optimization problem, but it is often unclear when multistart local search is sufficient and when population-based global search is worth the extra evaluations. We address this with a landscape-first design. Controlled QAOA experiments first identify two mechanisms that make local optimization unreliable: repeated use of the same small parameter set through multiple circuit layers, and competition between 2-local and 3-local cost terms. Simply increasing QAOA depth does not produce the same effect. We then test these mechanisms on eight previously unseen spin-glass instances, fresh $N=10$ and $N=12$ instances, and standard MaxCut, transverse-field Ising, and Heisenberg variational models. On all eight unseen spin glasses, at least one of the three tested adaptive differential-evolution variants achieves lower median error than both multistart BFGS and multistart Powell for the two difficult constructions; the independent-depth control does not. The parameter-reuse effect also transfers to MaxCut at both tested sizes, whereas the standard VQE models remain favorable to local search. Finally, a pre-benchmark landscape score combining random-start local-search outcomes and one-dimensional parameter-space slices is fixed from the original confirmation data before new optimizer outcomes are evaluated. It predicts whether the best of the three tested adaptive-DE variants outperforms MS-BFGS by more than one spectral-range percentage point on 50 new quantum objectives with 80--86\% condition-level accuracy across evaluation budgets. Within the tested families, the clearest signal for global-search usefulness is that random-start local search frequently ends in meaningfully inferior basins, rather than circuit depth or local curvature anisotropy. The results also motivate a broader ansatz-design question: whether some quantum-circuit complexity can be exchanged for harder classical optimization when robust global search is available.
\end{abstract}

\maketitle

\section{Introduction}

Variational quantum algorithms (VQAs) define a classical optimization problem through a parameterized quantum state,
\begin{equation}
f_{H,U}(\boldsymbol{\theta})
=
\langle 0|U^\dagger(\boldsymbol{\theta}) H U(\boldsymbol{\theta})|0\rangle .
\end{equation}
The variational quantum eigensolver (VQE) and the quantum approximate optimization algorithm (QAOA) are canonical examples \cite{Peruzzo2014,Farhi2014,Cerezo2021Review,Tilly2022}. Their practical performance depends on both the quantum model and the classical optimizer. The optimization landscape is therefore a property of the Hamiltonian--ansatz pair $(H,U)$ and its parameterization, not of the Hamiltonian alone.

A practical optimizer-selection question follows immediately: \emph{when is a strong local optimizer sufficient, and when is sustained global search useful?} Existing VQA studies usually answer this empirically after choosing a physical problem and ansatz. Barren-plateau theory characterizes regimes with exponentially suppressed gradients \cite{McClean2018,Cerezo2021Cost,Larocca2025}; separate results show that poor local minima can remain abundant even when gradients are finite \cite{AnschuetzKiani2022,Nemkov2025}; and VQA training can be computationally hard even for restricted models \cite{BittelKliesch2021}. QAOA studies have also reported reachability limits, local-minimum sensitivity, and nontrivial basin organization \cite{SackSerbyn2021,Akshay2021,BoyWales2024}. None of these properties alone determines whether a finite-budget local or population-based optimizer will be preferable.

Optimizer benchmarks show the same problem dependence. Differential Evolution (DE) \cite{StornPrice1997} can avoid local minima that trap conventional local methods on selected VQE instances \cite{Failde2023}; particle-swarm optimization and other population methods can also be effective \cite{Mei2024}; and broad VQE optimizer comparisons report substantial changes in ranking across models and metrics \cite{Jones2025, illesova2025qmetric, Illesova2025Statistical}. Recent noisy-VQA studies further show that sampling noise, ansatz choice, initialization, and parameter activity can all alter optimizer rankings \cite{bonet2023performance,BoyWales2024,Bezdek2025ClassicalOptimization,Novak2025Reliable, Novak2025NoisyLandscapes}. This motivates separating intrinsic exact-objective geometry from noise-induced difficulty.

Continuous black-box optimization takes a more explicit problem-feature view. BBOB \cite{hansen2010comparing} and CEC \cite{Novak2026CEC} suites deliberately construct multimodality, nonseparability, hybrid structure, and other features to expose solver strengths and weaknesses \cite{finck2010,awad2017}. Benchmark choice can materially change algorithm rankings \cite{Piotrowski2023}, and modern benchmarking guidance therefore emphasizes matched budgets, repeated runs, strong baselines, and nonparametric statistics \cite{LaTorre2021}. Exploratory landscape analysis and algorithm-footprint methods similarly seek features that explain algorithm success and failure \cite{Nikolikj2025,Cenikj2026,Novak2026Landscape}, although generalization of feature-based algorithm selection across problem families remains difficult \cite{Cenikj2025}.

\begin{figure*}[htpb]
\centering
\paperfig{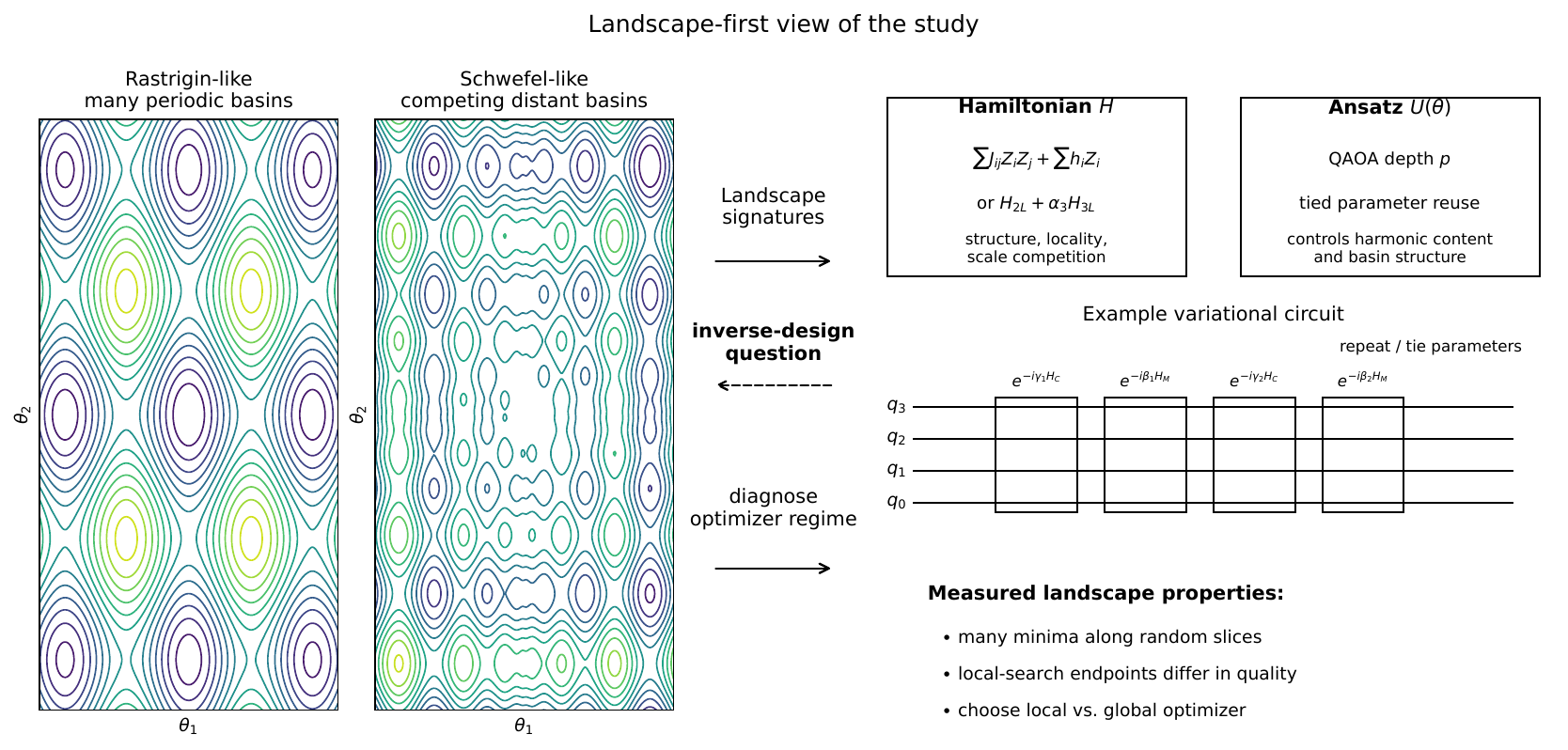}{0.95\textwidth}
\caption{Conceptual view of the landscape-first question. Classical multimodal functions provide familiar examples of landscapes with many competing basins. In a VQA, the objective landscape is instead induced jointly by the Hamiltonian $H$ and parameterized circuit $U(\boldsymbol{\theta})$. We vary these ingredients in controlled ways, measure the resulting basin structure and one-dimensional slice complexity, and ask whether those measurements indicate when multistart local search is sufficient or when sustained global evolutionary search is useful. The drawing is schematic; the classical contours are illustrative reference functions rather than data from the quantum experiments.}
\label{fig:intro_workflow}
\end{figure*}

The present work applies that logic to VQAs. Instead of sampling hundreds of unrelated Hamiltonian--ansatz pairs, we ask which controlled changes to $(H,U)$ create geometric motifs analogous to those that make classical functions such as Rastrigin and Schwefel difficult: repeated wells, competing basins, and local-search endpoints of substantially different quality. We then test whether those mechanisms transfer to new Hamiltonians and whether pre-benchmark landscape diagnostics can predict the optimizer regime before a full optimizer benchmark is run. Throughout, \emph{Rastrigin-like} and \emph{Schwefel-like} refer to these geometric analogies rather than functional equivalence.

This leads to two complementary parts of the study. The first is \emph{mechanistic}: controlled construction and confirmation identify changes that shift the finite-budget advantage from local to global search. The second is \emph{predictive}: the strongest mechanisms and negative controls are moved to unseen instances and standard VQA models, and a decision rule based on landscape diagnostics is frozen before new optimizer results are observed. The first part establishes what changes the landscape; the second tests whether that knowledge transfers.

The main questions are therefore:
\begin{enumerate}
\item Can circuit complexity be increased at fixed parameter dimension in a way that reliably reduces the effectiveness of multistart local search?
\item Can competing Hamiltonian structures create the same effect?
\item Which directly interpretable landscape measurements distinguish difficult from locally accessible objectives?
\item Do the identified mechanisms transfer to unseen spin-glass instances, different system sizes, and standard MaxCut/VQE models?
\item Can landscape measurements made before optimizer benchmarking predict when adaptive DE will provide a meaningful advantage?
\end{enumerate}

\section{Methods}

\subsection{Study overview}

The study has three phases, summarized in Table~\ref{tab:study_overview}. Phase I is exploratory and is used only to identify candidate mechanisms. Phase II is a controlled confirmation experiment. Phase III is an out-of-sample generalization and prospective-prediction experiment. This separation is important: the landscape score used in Phase III is calibrated only from Phase II and is fixed before the Phase III optimizer trajectories are examined.

\begin{table*}[htpb]
\centering
\caption{Study design. The phases have different roles: discovery, controlled confirmation, and out-of-sample validation.}
\label{tab:study_overview}
\maintablestyle
\begin{tabular}{@{}p{0.06\textwidth}p{0.17\textwidth}p{0.39\textwidth}p{0.32\textwidth}@{}}
\toprule
\textbf{Phase} & \textbf{Purpose} & \textbf{Main design} & \textbf{Role in inference} \\
\midrule
I & Mechanism discovery & One-run screen over QAOA depth, repeated parameter use, interaction locality, mixer/coupling heterogeneity, and VQE controls & Exploratory only; used to choose candidate mechanisms \\
II & Controlled confirmation & 18 conditions, exact objectives, 15 paired runs per primary condition, five optimizers, three function-evaluation (FE) checkpoints & Establish controlled mechanism effects and landscape--performance associations \\
III & Generalization and prediction & 50 new quantum conditions plus four classical controls, eight runs, six optimizers; eight unseen disorder instances, $N=10,12$ transfer, MaxCut, transverse-field Ising and Heisenberg VQE & Test transfer and evaluate predictions fixed before new optimizer outcomes \\
\bottomrule
\end{tabular}
\end{table*}

All quantum objectives are evaluated by exact statevector simulation. For a quantum condition, optimization performance is reported as normalized physical error
\begin{equation}
\epsilon_{\rm phys}(\boldsymbol{\theta})
=
\frac{f(\boldsymbol{\theta})-E_0}{E_{\max}-E_0},
\label{eq:phys}
\end{equation}
where $E_0$ and $E_{\max}$ are the spectral extrema of the Hamiltonian. Thus $\epsilon_{\rm phys}=0$ corresponds to the exact ground energy and $\epsilon_{\rm phys}=1$ to the top of the Hamiltonian spectrum. The normalization is an affine rescaling within each condition, so it does not change optimizer rankings. It only expresses an energy error relative to that Hamiltonian's spectral width. Diagonal spectra are enumerated exactly and non-diagonal extrema are obtained with sparse eigensolvers.

Raw normalized-error differences are naturally small decimals. For readability, optimizer advantages and local-search endpoint gaps in the main-text figures and tables are therefore reported as \emph{spectral-range percentage points (pp)}: a difference of $0.04$ in Eq.~\eqref{eq:phys} is shown as $4.0$. Thus the plotted number is a signed percentage-point difference, not a percentage constrained to $0$--$100$; $+4$ means that DE lowers the normalized error by $0.04$, while $-4$ means the local baseline is better by the same amount. Appendix tables retain the underlying normalized values where useful for direct reproduction.

\subsection{Optimizer panel and rationale}

The optimizer panel is designed to answer a local-versus-global search question rather than to survey every optimizer used in VQAs. We therefore use strong representatives of qualitatively different search mechanisms. Multistart BFGS (MS-BFGS) is the primary local baseline because quasi-Newton descent is efficient on smooth exact objectives and, when restarted from independent points, is substantially stronger than a single local run \cite{NocedalWright2006}. Phase III adds multistart Powell (MS-Powell), a derivative-free direction-set local method, to test whether a global-search advantage is specific to BFGS or to numerical-gradient information.

MS-CMA-ES provides a population-based comparison that is not based on differential evolution: it adapts a multivariate search distribution and can follow anisotropic directions without using derivatives \cite{HansenOstermeier2001}. The global-search family of primary interest contains iL-SHADE, jSO, and L-SRTDE. These are mature adaptive Differential Evolution (DE) variants from the CEC single-objective optimization lineage. They share population-based mutation and greedy selection but differ in how control parameters, population size, and exploitation pressure are adapted \cite{StornPrice1997,TanabeFukunaga2013,TanabeFukunaga2014,Brest2016iLSHADE,Brest2017jSO,Stanovov2024LSRTDE}. Using three adaptive-DE variants reduces the chance that a result is peculiar to one DE implementation.

\begin{table}[htpb]
\centering
\caption{Role and inclusion rationale for the optimizer benchmark panel.}
\maintablestyle
\label{tab:optimizer_rationale}
\begin{tabular}{@{}p{0.21\columnwidth}p{0.26\columnwidth}p{0.44\columnwidth}@{}}
\toprule
\textbf{Method} & \textbf{Class / Paradigm} & \textbf{Benchmark rationale} \\
\midrule
MS-BFGS   & Local, quasi-Newton   & Smooth local baseline; tests basin accessibility via restarts. \\
MS-Powell & Local, derivative-free  & Direction-set local baseline; controls for gradient artifacts. \\
MS-CMA-ES & Evolution strategy  & Population baseline with adaptive Gaussian sampling (non-DE mechanism). \\
iL-SHADE  & Adaptive DE       & Standard reference with success-history adaptation and linear population-size reduction. \\
jSO       & Adaptive DE       & Tuned reference with alternative parameter adaptation schedules. \\
L-SRTDE   & Adaptive DE       & Success-rate-based adaptation reference. \\
\bottomrule
\end{tabular}
\end{table}

All methods receive the same condition-specific FE budget. One exact objective evaluation at one parameter vector counts as one FE, including evaluations used internally for numerical gradients or line searches. All optimizer coordinates use the common bounded box $[-\pi,\pi)^D$; population methods are explicitly bounded to that box and local-method evaluations outside it are mapped back coordinatewise by the common wrapper. This is a numerical boundary convention used consistently across methods, not a claim that every heterogeneous generator is exactly $2\pi$-periodic. In particular, the incommensurate-mixer and Gaussian-coupling Phase-II controls should be interpreted as the wrapped benchmark objectives induced by this convention rather than unrestricted physical parameterizations. The original 14-condition calibration, fixed before Phase III, is retained as the primary prospective analysis. Replicate runs use the same integer seed label across optimizers; this provides a common replicate block but not a common random-number path across different algorithms. The study deliberately uses exact deterministic objectives, so noise-specialized stochastic-gradient methods are outside the scope of the present geometry-focused comparison.

\paragraph{Quantum models and circuit conventions.}
For the QAOA conditions, $H_M=\sum_iX_i$ is the transverse-field mixer and $H_C$ is the diagonal cost Hamiltonian. Writing $\gamma_\ell$ and $\beta_\ell$ for the effective angles used by the simulator,
\begin{align}
|\psi(\boldsymbol\gamma,\boldsymbol\beta)\rangle &= U_p\cdots U_1|+\rangle^{\otimes N},
& U_\ell&=e^{-i\beta_\ell H_M}e^{-i\gamma_\ell H_C},\\
|\psi_r\rangle &= U_3^rU_2^rU_1^r|+\rangle^{\otimes N}.&&
\end{align}
Thus tied reuse repeats each individual layer before moving to the next one; it does not repeat the complete $p=3$ block. The Phase-II reference cost is the saved frustrated Ising spin glass $H_C=\sum_{(i,j)\in E}J_{ij}Z_iZ_j+\sum_i h_iZ_i$, where $E$ is a ring plus opposite-node chords (degree three for even $N$), $J_{ij}\in\{-1,+1\}$, and $h_i\sim\mathcal U[-0.2,0.2]$. The reference instance was selected from 50 generated candidates by its number of one-spin-flip local minima. Phase-III spin glasses instead use fresh, unselected connected 3-regular graphs with independent $\pm1$ couplings and fields from the same interval. Unweighted MaxCut uses the minimization-equivalent $H_{\rm MC}=\tfrac12\sum_{(i,j)\in E}Z_iZ_j$, with the irrelevant additive constant omitted. The hardware-efficient VQE ansatz applies $R_y$ rotations on all qubits, followed by the nearest-neighbor CNOT chain $0\!\to\!1,1\!\to\!2,\ldots,N\!-\!2\!\to\!N\!-\!1$ for each of $L$ entangling layers, and ends with a final $R_y$ layer.

\subsection{Phase II: controlled construction mechanisms}

The confirmation panel contains 14 quantum conditions and four classical controls. Standard QAOA uses the Ising spin-glass cost defined above at depth $p=3$, hence six parameters. Two construction mechanisms are central.

First, \emph{tied parameter reuse} repeats each of the three $p=3$ cost--mixer layers $r$ times while tying that layer's $(\gamma,\beta)$ pair across its repetitions. Reuse factors $r=2,4,6$ therefore execute 6, 12, and 18 alternating cost--mixer layers while keeping the parameter dimension fixed at $D=6$. The independent-depth control instead uses $p=6$ with six independently parameterized layers and $D=12$. Thus $r=2$ and independent $p=6$ are matched in alternating-layer count but differ in parameter tying.

Second, the \emph{mixed-locality} construction combines 2-local and 3-local diagonal cost terms,
\begin{equation}
H(\alpha_3)=H_{2{\rm L}}+\alpha_3 H_{3{\rm L}},
\qquad
\alpha_3\in\{0.5,1,2\},
\end{equation}
with the same $p=3$ QAOA parameterization. The 3-local component uses $1.5N$ distinct randomly selected triples with raw $\pm1$ coefficients and is rescaled so that its coefficient-vector $\ell_2$ norm matches that of the base up-to-2-local cost before multiplication by $\alpha_3$. Hence $\alpha_3=1$ represents matched coefficient-norm scale. Additional controls vary mixer weights and coupling magnitudes, and a hardware-efficient VQE condition provides a non-QAOA comparison. Rastrigin and Schwefel at $D=6$ and $D=10$ are included only as classical multimodal sanity checks.

Phase II uses MS-BFGS, MS-CMA-ES, iL-SHADE, L-SRTDE, and jSO from the panel above. Each primary condition has 15 paired runs with common replicate seeds. A single trajectory is continued to 30,000 function evaluations (FEs) and sampled at 5,000, 15,000, and 30,000 FEs. The multistart methods continue launching fresh local searches until the same total FE budget is exhausted; they are therefore budget-matched local-search strategies rather than single-start baselines.

\subsection{Phase III: unseen instances and standard VQA models}

Phase III keeps the strongest Phase-II mechanisms---tied reuse $r=6$ and mixed locality with $\alpha_3=1$---together with standard $p=3$ QAOA and the independent-depth $p=6$ negative control.

The main instance-level validation uses eight previously unseen $N=10$ random connected 3-regular Ising spin glasses. Graph topology, $\pm J$ couplings, and longitudinal fields are generated independently for each instance; no instance is selected by landscape difficulty or optimizer performance. Each physical instance is evaluated with the same four QAOA variants, giving a paired mechanism comparison across eight Hamiltonians.

A separate size-transfer set uses fresh random 3-regular spin-glass instances at $N=10$ and $N=12$, again with the four QAOA variants. Transfer beyond the spin-glass family is tested using unweighted MaxCut QAOA on unselected connected 3-regular graphs at $N=10$ and $N=12$, with standard $p=3$, independent $p=6$, and tied reuse $r=6$. Standard VQE baselines use a periodic transverse-field Ising (TFIM) model,
\begin{equation}
    H_{\rm TFIM}=-\sum_i Z_iZ_{i+1}-\sum_i X_i,
\end{equation}
and an open antiferromagnetic Heisenberg chain,

\begin{equation}
   H_{\rm Heis}=\sum_i (X_iX_{i+1}+Y_iY_{i+1}+Z_iZ_{i+1}),
\end{equation}
with $R_y$--CNOT hardware-efficient ansatzes (HEAs) of one and two entangling layers. These models test whether the proposed diagnostics merely label all nontrivial quantum objectives as globally difficult.

The final Phase-III panel contains 50 new quantum conditions and four classical Rastrigin/Schwefel controls. Each condition uses eight runs and the same 30,000-FE horizon. MS-Powell is added as a second multistart local baseline, giving six optimizers in total: MS-BFGS, MS-Powell, MS-CMA-ES, iL-SHADE, L-SRTDE, and jSO.

A planned $N=14$ extension was stopped when exact-statevector runtime became substantially larger than at $N=10$ and $N=12$. Its partial optimizer runs are excluded from every analysis reported here. The exclusion was made for computational cost, after the landscape-based predictions had been frozen, and not on the basis of optimizer outcomes.

\subsection{Local-search and landscape diagnostics}

We use standard optimization terminology wherever possible. A \emph{random-start local minimization} means a BFGS run initialized uniformly in the bounded coordinate box and continued under a fixed per-start local-search budget. In energy-landscape terminology such a run is often called a quench; here we avoid that term because the optimization interpretation is more direct.

For $m$ random-start local minimizations with final normalized errors $e_1,\ldots,e_m$ and best observed endpoint $e_{\min}$, we use three basin-accessibility measurements:
\begin{align}
p_{\rm near}
&=
\frac{1}{m}\sum_{i=1}^m
\mathbf{1}[e_i-e_{\min}\le 0.01],
\\
g_{\rm med}
&=
\operatorname{median}_i(e_i-e_{\min}),
\\
p_{\rm poor}
&=
\frac{1}{m}\sum_{i=1}^m
\mathbf{1}[e_i-e_{\min}>0.05].
\end{align}
Here $p_{\rm near}$ is the fraction of local starts that reach a near-best endpoint, $g_{\rm med}$ is the typical quality loss of a local-search endpoint relative to the best observed endpoint, and $p_{\rm poor}$ is the fraction of starts that terminate substantially above the best endpoint. These quantities operationally measure the capture frequency and quality distribution of local-search outcomes without requiring enumeration of all stationary points.

Parameter-space corrugation is probed using random one-dimensional wrapped line segments in the common coordinate box. For each segment we count the number of sampled local minima and compute the normalized Fourier entropy of the objective profile. The first quantity measures how many oscillatory wells a typical direction crosses; the second measures how broadly the variation is distributed over Fourier frequencies. Finite-difference gradients, barriers, distance--energy relations, and Hessians are also recorded. Phase II uses 128 random-start BFGS minimizations per selected condition; Phase III uses 64. These diagnostics are computed once per condition and are not charged against any optimizer's benchmark FE budget; they are used as a pre-benchmark decision probe rather than as a claim of net FE savings. For a representative Phase-III $D=6$ QAOA condition, the five features entering the predictor require at most 25,376 objective evaluations: $64\times300=19{,}200$ from the local searches and $32\times193=6{,}176$ from the line scans. This is comparable to one 30,000-FE optimizer trajectory and about 1.8\% of the full $6\times8\times30{,}000=1.44$ million evaluations used by the condition-level optimizer benchmark.

\begin{table*}[htpb]
\centering
\caption{Landscape diagnostics, operational definitions, and geometric interpretations.}
\label{tab:diagnostics_plain}
\maintablestyle
\begin{tabular}{@{}p{0.18\textwidth}p{0.34\textwidth}p{0.42\textwidth}@{}}
\toprule
\textbf{Quantity} & \textbf{Operational Definition} & \textbf{Geometric Interpretation} \\
\midrule
$p_{\rm near}$ &
  Fraction of random BFGS starts ending within $0.01$ normalized error of the best sampled local-search endpoint. &
  Empirical capture frequency of near-best local-search outcomes; high values indicate that good endpoints are easy to reach from random starts. \\

$g_{\rm med}$ &
  Median gap between local-search endpoints and the best sampled local-search endpoint. &
  Measures endpoint-quality dispersion; high values imply substantial loss when local search reaches an inferior basin. \\

$p_{\rm poor}$ &
  Fraction of starts terminating $>0.05$ normalized error above the best sampled local-search endpoint. &
  Empirical frequency of substantially inferior local-search outcomes. \\

Minima per line &
  Mean number of sampled local minima along random wrapped 1D line segments. &
  Directional multimodality (corrugation density); counts crossed ridges without capturing relative well depths. \\

Fourier entropy (1D) &
  Spectral entropy of random 1D objective profiles. &
  Multiscale structure; low values imply dominant wavelengths, while high values reflect irregular, multi-frequency roughness. \\

Positive Hessian eigenvalue ratio &
  $\max(\lambda_+)/\min(\lambda_+)$ over retained positive Hessian eigenvalues at the best sampled local-search endpoint; numerically negligible modes are excluded. &
  Local positive-curvature anisotropy near one endpoint; does not describe distant basin quality or accessibility. \\
\bottomrule
\end{tabular}
\end{table*}

An eigenvalue is treated as active when $|\lambda_i|>\max(10^{-9},10^{-6}\max_j|\lambda_j|)$; the reported positive Hessian eigenvalue ratio is computed only over active positive modes.

A useful geometric distinction is between \emph{corrugation} and \emph{basin-quality heterogeneity}. A surface can contain many wells but remain easy for multistart local optimization when the wells have similar depths or when a large fraction of random starts reaches good basins. Global search becomes more valuable when the surface is multimodal and the basins reached by local search differ substantially in quality. This distinction is central below. Independent QAOA depth mainly increases corrugation, whereas tied parameter reuse and mixed locality increase both corrugation and the spread in basin quality.

Representative empirical slices of the exact quantum objectives are shown directly in the Results (Section~\ref{sec:empirical_slices}), while Appendix~\ref{app:landscape_geometry} provides complementary schematics explaining what the diagnostics mean geometrically.

\subsection{Evaluation protocol and statistical comparisons}

For compact family-level summaries, let
\begin{equation}
\begin{aligned}
\DEadvB(B)={}&\operatorname{median}\epsilon_{\rm MS-BFGS}(B)\\
&-\min_{a\in\mathcal{D}}\operatorname{median}\epsilon_a(B),
\end{aligned}
\label{eq:de_bfgs}
\end{equation}
where $\mathcal{D}=\{\text{iL-SHADE},\text{L-SRTDE},\text{jSO}\}$. A positive value means that at least one adaptive-DE method has a lower median error than MS-BFGS. In Phase III we also report
\begin{equation}
\begin{aligned}
\DEadvL(B)={}&\min_{\ell\in\{\mathrm{BFGS},\mathrm{Powell}\}}
\operatorname{median}\epsilon_\ell(B)\\
&-\min_{a\in\mathcal{D}}\operatorname{median}\epsilon_a(B),
\end{aligned}
\label{eq:de_local}
\end{equation}
which compares the adaptive-DE family with the better of the two local baselines. These family-level differences are descriptive because the best DE method is selected after observing the medians. Inferential comparisons use named optimizers: Friedman omnibus tests followed by paired Wilcoxon tests and Holm correction within each condition and checkpoint. We additionally report paired win fractions, rank-biserial effects where applicable, and bootstrap confidence intervals. Run-level pairing uses the common replicate index as a blocking label; because different algorithms consume random numbers differently, it is not interpreted as a common-random-number experiment. In the eight-instance holdout experiment, the Hamiltonian instance is the replication unit for mechanism-level tests: variants are paired within the same physical instance and compared by two-sided paired Wilcoxon tests, with Holm correction across the three prespecified variant-versus-standard contrasts.

For readability, Eqs.~\eqref{eq:de_bfgs}--\eqref{eq:de_local} are defined in normalized-error units, but main-text tables and figures multiply these differences by 100 and display them as percentage points of the Hamiltonian spectral range. A positive value always means that adaptive DE attains the lower median error.

\subsection{Prospective prediction from landscape diagnostics}

Phase II shows which landscape quantities are associated with a shift from local-search advantage to global-search advantage, but those within-panel correlations are not an out-of-sample test. Phase III therefore uses a deliberately simple predictor fixed before any new optimizer results are evaluated.

Five Phase-II measurements are used: median local-search endpoint gap, poor-endpoint fraction, minima per random line, Fourier entropy, and the negative of the near-best local-search rate. Each feature is robustly standardized by its Phase-II median and interquartile range, clipped to $[-3,3]$, and the five standardized values are averaged. The resulting \emph{landscape difficulty score} is larger when random local search tends to end in unequal basins and when random slices are more multimodal.

At each FE checkpoint, a threshold for this score is chosen using only the 14 Phase-II primary quantum conditions by maximizing balanced accuracy; if several thresholds tie, the least extreme threshold (closest to zero) is used. The binary target is whether Eq.~\eqref{eq:de_bfgs} exceeds 0.01 normalized error, i.e. whether the best median among the three tested adaptive-DE variants improves on MS-BFGS by more than one spectral-range percentage point. The prediction therefore concerns the existence of a substantive advantage within the tested adaptive-DE panel, not selection of a particular DE variant. The score definition, threshold, calibration inputs, and all predictions for the new panel are written to disk and SHA-256 hashed before the Phase-III optimizer outcomes are evaluated. Prediction metrics are reported at the condition level on the retained 50 new quantum conditions. Because several conditions are variants of the same Hamiltonian, dependence-aware uncertainty is checked by a grouped bootstrap that resamples the 14 underlying base instances rather than individual conditions.

Because two Phase-II controls---the incommensurate mixer and Gaussian couplings---use heterogeneous generators under the common wrapping convention, we also perform a post-hoc sensitivity analysis in which the same score is recalibrated after excluding those two controls. This sensitivity analysis does not alter the frozen primary predictions and is used only to assess whether the primary result depends materially on those two wrapped benchmark objectives.

\section{Results}

\subsection{Controlled construction establishes two difficult-landscape mechanisms}

Phase II first asks whether a shift from local-search advantage to global-search advantage can be produced by a controlled change rather than discovered accidentally in a large benchmark set. Tied parameter reuse gives the clearest trend. With the QAOA parameter dimension fixed at $D=6$, increasing reuse from $r=1$ to $r=6$ raises the median local-search endpoint gap from 5.2 to 29.0 percentage points of the spectral range and the average number of minima on a random line from 3.73 to 13.64. At the same time, the best adaptive-DE median becomes progressively better than the MS-BFGS median (Table~\ref{tab:tied}). The effect persists to 30,000 FEs.

\begin{table*}[htpb]
\centering
\caption{Controlled tied-reuse QAOA experiment at fixed $D=6$. Landscape accessibility and adaptive-DE advantage are reported on directly interpretable scales; positive advantage means lower median error than MS-BFGS.}
\label{tab:tied}
\maintablestyle
\begin{tabular}{@{}crrrrrr@{}}
\toprule
\textbf{Reuse} &
\multicolumn{1}{c}{\mhead{Near-best\\outcomes (\%)}} &
\multicolumn{1}{c}{\mhead{Median endpoint gap\\(spectral-range pp)}} &
\multicolumn{1}{c}{\mhead{Minima\\per line}} &
\multicolumn{1}{c}{\mhead{Fourier\\entropy}} &
\multicolumn{1}{c}{\mhead{DE advantage\\5k (pp)}} &
\multicolumn{1}{c}{\mhead{DE advantage\\30k (pp)}} \\
\midrule
1 & 1.6 & 5.2 & 3.734 & 0.284 & 0.3 & 0.5 \\
2 & 0.8 & 9.4 & 5.312 & 0.332 & 2.4 & 1.8 \\
4 & 0.8 & 17.3 & 10.141 & 0.420 & 4.2 & 1.8 \\
6 & 1.6 & 29.0 & 13.641 & 0.446 & 7.7 & 4.0 \\
\bottomrule
\end{tabular}
\end{table*}

\begin{figure}[htpb]
\centering
\paperfig{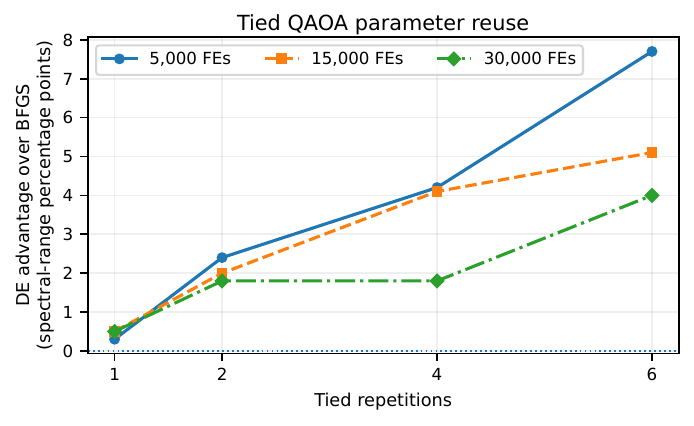}{0.98\linewidth}
\caption{Controlled tied-reuse experiment. Repeating each of the three QAOA cost--mixer layers while tying its angle pair progressively increases the finite-budget advantage of adaptive DE over multistart BFGS. The vertical axis is expressed in spectral-range percentage points; e.g., $4$ means a normalized-error difference of $0.04$, while negative values indicate the opposite sign of advantage.}
\label{fig:tied}
\end{figure}

Simply making QAOA deeper does not have the same effect. The independent $p=6$ control has twice as many parameters and a much larger positive Hessian eigenvalue ratio, yet MS-BFGS is better at 5,000 FEs and the difference is approximately zero by 30,000 FEs. The matched-layer comparison is especially informative: tied $r=2$ and independent $p=6$ both execute six alternating cost--mixer layers, but only the tied construction shows a positive adaptive-DE advantage. Thus layer count and local curvature anisotropy are not sufficient explanations; how parameters are reused also matters.

\begin{figure}[htpb]
\centering
\paperfig{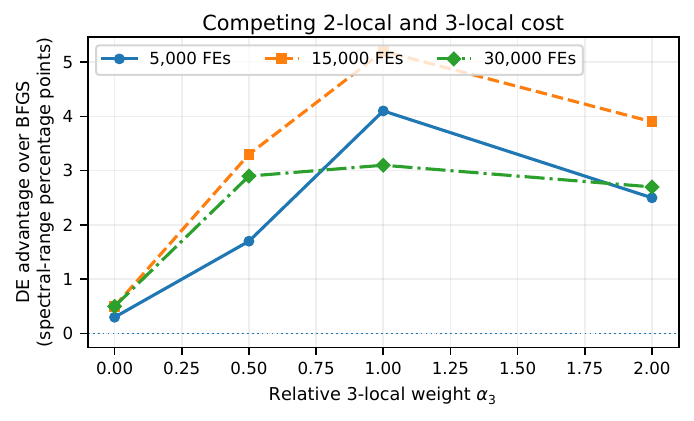}{0.98\linewidth}
\caption{Controlled mixed-locality experiment. The largest early global-search advantage occurs when 2-local and 3-local structures have comparable coefficient-norm scale. The vertical axis gives the adaptive-DE improvement over MS-BFGS in spectral-range percentage points.}
\label{fig:locality}
\end{figure}

The second reproducible mechanism is competition between 2-local and 3-local cost terms. All three nonzero 3-local weights give a positive adaptive-DE advantage at every checkpoint, with the strongest early effect when the two structures have comparable coefficient-norm scale (Figure~\ref{fig:locality}). At 5,000 FEs the advantages are 1.7, 4.1, and 2.5 spectral-range percentage points for $\alpha_3=0.5,1,2$, respectively; at 30,000 FEs they remain 2.9, 3.1, and 2.7 points. The standard $\alpha_3=0$ case stays near zero (0.3--0.5 points). This non-monotone dependence argues against ``higher locality'' as a scalar difficulty measure; competing structures are more important than simply adding stronger higher-order interactions. The complete checkpoint values are also retained in Appendix Table~\ref{tab:all_phase2}.

The landscape measurements explain the difference between these positive and negative controls. Across the 14 Phase-II quantum conditions, the strongest descriptive associations with adaptive-DE advantage are median local-search endpoint gap ($\rho=0.846$ at 30k), poor-endpoint fraction ($\rho=0.811$), minima per random line ($\rho=0.741$), and line Fourier entropy ($\rho=0.732$). The positive Hessian eigenvalue ratio is weaker ($\rho=0.499$). These correlations are descriptive because the Phase-II conditions share backbones and sweep structure; their role is to motivate the independent test below rather than to establish a universal predictor.

The next subsection makes this mechanism claim visual on a matched unseen spin-glass instance: the independent-depth $p=6$ control mainly adds oscillation, whereas tied reuse and mixed locality create either many repeated wells or more clearly separated competing basins---the qualitative geometries most closely analogous to the Rastrigin-like and Schwefel-like behaviors discussed in the paper.

\begin{table}[htpb]
\centering
\caption{Phase-II descriptive Spearman correlations with adaptive-DE advantage.}
\label{tab:corr}
\maintablestyle
\begin{tabular}{@{}lrr@{}}
\toprule
Landscape measurement & 5k & 30k\\
\midrule
median local-search endpoint gap & 0.758 & 0.846 \\
poor-endpoint fraction $>0.05$ & 0.783 & 0.811 \\
near-best local-search rate $p_{\rm near}$ & -0.575 & -0.495 \\
minima per random line & 0.771 & 0.741 \\
line Fourier entropy & 0.771 & 0.732 \\
positive Hessian eigenvalue ratio & 0.389 & 0.499 \\
\bottomrule
\end{tabular}
\end{table}

Rastrigin and Schwefel provide a sanity check that the optimizer implementations show the expected local/global reversal against MS-BFGS. Their absolute normalized errors are not directly comparable with the quantum objectives. The Phase-III Powell baseline also shows why these functions should remain sanity controls rather than definitions of global difficulty: in low dimension, Powell can solve some of them competitively.

\subsection{The mechanism-induced quantum landscapes can be seen directly}
\label{sec:empirical_slices}

\begin{figure*}[htpb]
\centering
\paperfig{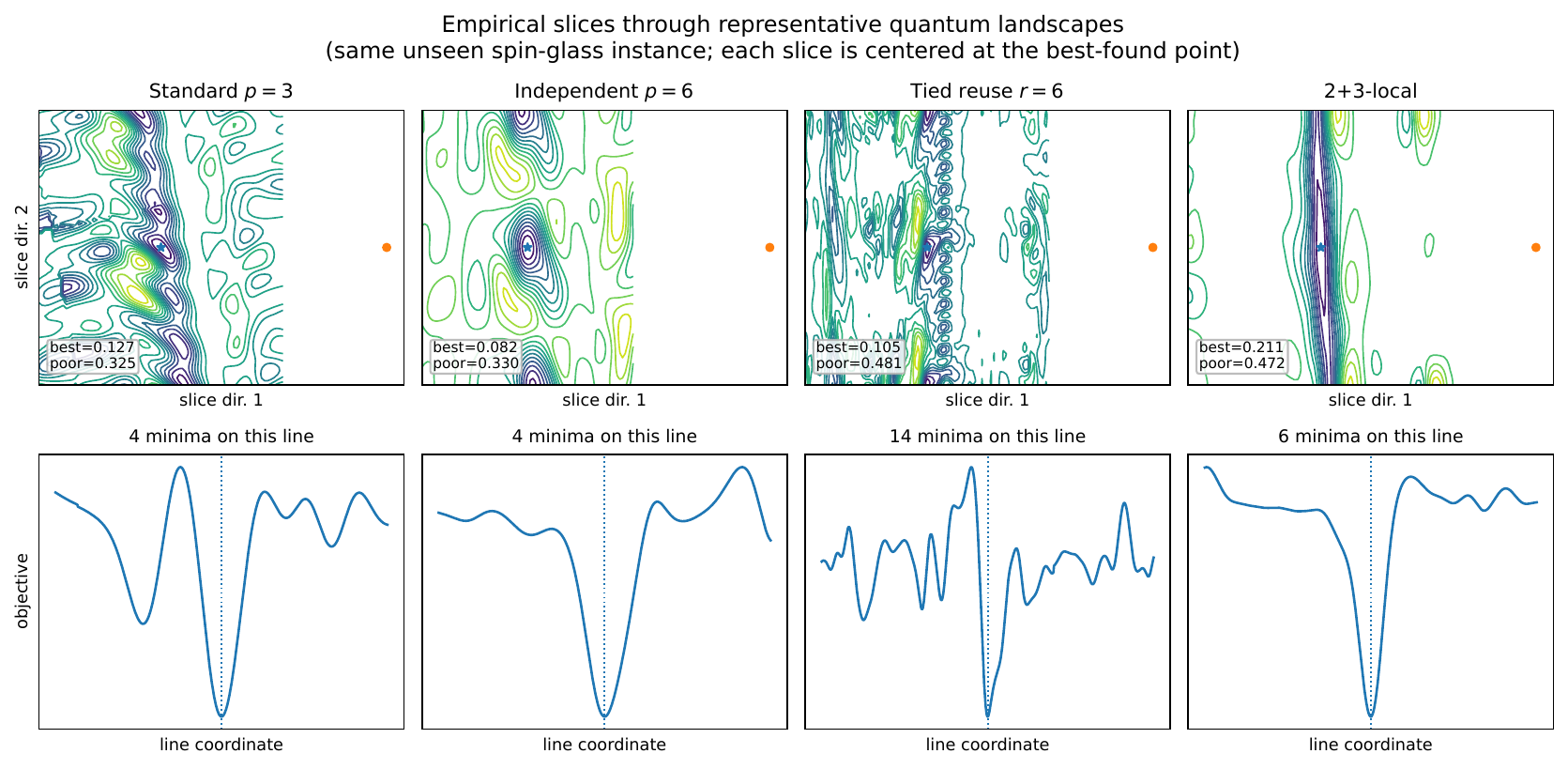}{0.95\textwidth}
\caption{Empirical slices through four exact quantum objectives built on the same unseen $N=10$ spin-glass backbone. Top: contour plots of two-dimensional slices centered at the best-found point; the star marks that best point and the circle marks a poorer local-search endpoint used to define the principal slice direction. Bottom: the corresponding one-dimensional slices along that direction. Standard $p=3$ shows only a few wells along the displayed line. Independent $p=6$ increases oscillatory structure but does not create equally severe basin separation. Tied reuse $r=6$ produces a much more Rastrigin-like line slice with many repeated minima, whereas the mixed-locality construction creates a narrow good basin separated from clearly worse alternatives, visually closer to a Schwefel-like competing-basin picture.}
\label{fig:empirical_quantum_slices}
\end{figure*}

Because the central claim is that specific Hamiltonian--ansatz mechanisms generate quantum landscapes with Rastrigin-like repeated-well structure or Schwefel-like competing basins, these geometric signatures should be visible rather than inferred only from scalar summaries. Figure~\ref{fig:empirical_quantum_slices} therefore plots a representative unseen $N=10$ spin-glass instance from the Phase-III holdout panel under four matched QAOA constructions: standard $p=3$, independent-depth $p=6$, tied reuse $r=6$, and the mixed-locality construction.

For each condition, we first locate a high-quality point using differential evolution and then collect many multistart BFGS endpoints. The displayed two-dimensional plane is spanned by the direction from the best-found point to a distinctly poorer local endpoint and one orthogonal companion direction; the one-dimensional slice is taken along that same good-to-bad direction. These plots are illustrative views through the full $D$-dimensional objective, not complete representations of the landscape.

These empirical slices support the distinction emphasized in the main text. Independent depth can increase corrugation without making local search systematically unreliable. By contrast, tied reuse combines a dense oscillatory structure with many competing minima along an informative line, and mixed locality creates a sharper separation between a good basin and poorer alternatives. The figure is therefore a visual companion to the quantitative results: it shows representative geometry, while the diagnostic and holdout statistics below establish that these effects are systematic rather than artifacts of a chosen slice.

\begin{figure*}[htpb]
\centering
\paperfig{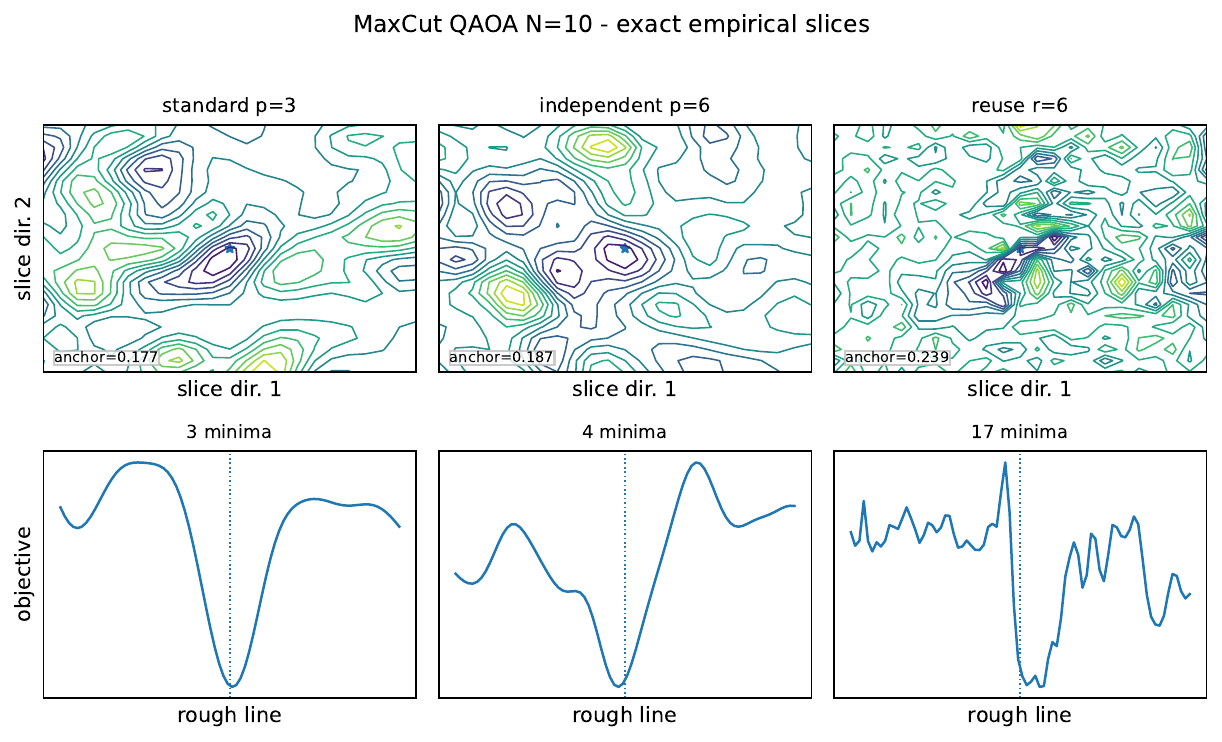}{0.75\textwidth}
\caption{High-corrugation empirical scan for a standard $N=10$ MaxCut QAOA instance. Among six random directions through a low-energy anchor, the line with the most local minima is shown together with an orthogonal two-dimensional slice. Standard $p=3$ and independent $p=6$ retain only a few wells, whereas tied reuse $r=6$ produces a dense repeated-well structure.}
\label{fig:empirical_maxcut}
\end{figure*}

The visual pattern is not limited to the spin-glass example. For the broader gallery we use a lighter high-corrugation scan: from a low-energy anchor, six random directions are evaluated and the line with the largest number of minima is retained; an orthogonal companion direction defines the contour plane. Figure~\ref{fig:empirical_maxcut} applies this scan to standard $N=10$ MaxCut QAOA. Standard $p=3$ and independent $p=6$ show only a few wells on the selected line, whereas tied reuse produces a dense repeated-well structure. Figure~\ref{fig:empirical_vqe} provides the complementary negative control: the TFIM and Heisenberg VQE models remain smooth and essentially unimodal even under the same deliberately rough-line selection. These scans are illustrative rather than inferential; the instance-level statistics below establish generality.

\subsection{The mechanisms replicate on eight unseen Hamiltonians}

\begin{figure}[htpb]
\centering
\paperfig{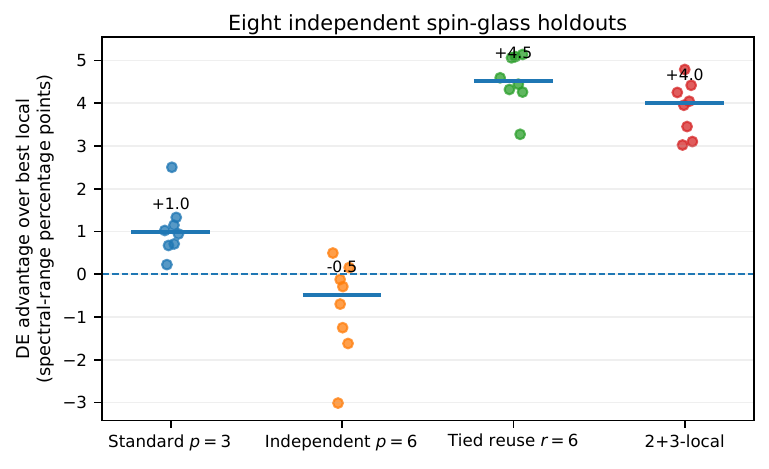}{0.98\linewidth}
\caption{Out-of-sample mechanism test on eight unseen $N=10$ random 3-regular spin glasses at 30,000 FEs. Each point is one independently generated Hamiltonian. Positive values mean that the descriptively best adaptive-DE median is lower than the better of the two multistart local baselines. Values are reported in spectral-range percentage points (so $+5$ means a normalized-error improvement of $0.05$); horizontal segments mark medians.}
\label{fig:holdout}
\end{figure}

The strongest test of the construction claim is the instance-level holdout experiment. Figure~\ref{fig:holdout} shows the adaptive-DE advantage over the \emph{better} of MS-BFGS and MS-Powell on eight independently generated spin glasses. Tied reuse and mixed locality are positive on all eight instances. Independent $p=6$ depth is positive on only two of eight.

\begin{table}[htpb]
\centering
\caption{Out-of-sample mechanism summary at 30,000 FEs across eight unseen spin-glass instances. Advantages are reported in spectral-range percentage points. ``DE wins'' counts instances for which the adaptive-DE advantage over the better local baseline is positive.}
\label{tab:holdout}
\maintablestyle
\begin{tabular}{@{}lrrr@{}}
\toprule
QAOA variant & vs. BFGS & vs. better local & DE wins\\
\midrule
standard $p=3$ & 0.98 & 0.98 & 8/8\\
independent $p=6$ & -0.49 & -0.49 & 2/8\\
tied reuse $r=6$ & 4.99 & 4.52 & 8/8\\
2+3-local, $\alpha_3=1$ & 4.00 & 4.00 & 8/8\\
\bottomrule
\end{tabular}
\end{table}

The paired nature of the holdout design allows a direct mechanism comparison within the same physical instance. Relative to standard $p=3$, tied reuse increases the DE-versus-BFGS advantage by a median 4.06 spectral-range percentage points, mixed locality by 3.06 points, whereas independent $p=6$ decreases it by 1.78 points. All eight instance-level differences have the same sign for each of these three contrasts; the two-sided paired Wilcoxon tests give raw $p=0.0078$ in each case and Holm-adjusted $p=0.0234$ across the three prespecified contrasts. When DE is compared with the better of BFGS and Powell, the corresponding median increases are 3.46 and 2.91 points for tied reuse and mixed locality, while the $p=6$ change remains $-1.78$ points.

The landscape measurements show why the independent-depth control is important. Independent $p=6$ does increase one-dimensional complexity, but it does not systematically worsen the quality distribution of local-search endpoints. Tied reuse and mixed locality increase both kinds of difficulty (Table~\ref{tab:holdout_landscape}). In particular, the median endpoint gap rises from about 5\% of the spectral range for standard QAOA to 20\% for tied reuse and 17\% for mixed locality, while roughly 90\% of random local starts end more than five spectral-range percentage points above the best local-search endpoint.

\begin{table}[t]
\centering
\caption{Median landscape measurements across the eight unseen spin-glass instances. Gap is in spectral-range percentage points.}
\label{tab:holdout_landscape}
\scriptsize
\renewcommand{\arraystretch}{1.05}
\setlength{\tabcolsep}{2.2pt}
\begin{tabular}{@{}lrrrrr@{}}
\toprule
Variant & Gap & Poor (\%) & Near-best (\%) & Min./line & Entropy \\
\midrule
standard $p=3$ & 5.0 & 49.2 & 3.9 & 3.75 & 0.280\\
independent $p=6$ & 5.4 & 50.8 & 5.5 & 5.48 & 0.360\\
tied reuse $r=6$ & 20.4 & 90.6 & 2.3 & 13.38 & 0.433\\
2+3-local & 17.1 & 89.8 & 2.3 & 5.55 & 0.327\\
\bottomrule
\end{tabular}
\end{table}

Thus ``more minima'' is not by itself the relevant criterion. Independent depth adds minima and spectral complexity but leaves local basin quality relatively benign; the two successful construction mechanisms create many local-search outcomes that are genuinely worse than the best accessible basin.

\subsection{Transfer to size scaling, MaxCut, and standard VQE models}

A fresh spin-glass size-transfer set reproduces the same qualitative ordering at both $N=10$ and $N=12$: tied reuse and mixed locality favor adaptive DE, independent $p=6$ favors the local baselines, and standard $p=3$ shows only a small advantage (Table~\ref{tab:transfer}).

More importantly, tied reuse transfers to a standard MaxCut QAOA setting. Standard $p=3$ MaxCut is essentially tied between local and DE search at 30,000 FEs, while independent $p=6$ favors local search. Reusing the same $p=3$ parameters six times changes the result: adaptive DE improves on the better local median by 1.68 spectral-range percentage points at $N=10$ and 2.77 points at $N=12$. The corresponding geometry changes are large. At $N=10$, the median local-search endpoint gap rises from 2.8\% to 28.1\% of the spectral range and the poor-endpoint fraction from 10.9\% to 90.6\%; the average number of minima per random line rises from 2.19 to 9.31. At $N=12$, the same quantities change from 2.1\% to 12.0\%, 4.7\% to 75.0\%, and 2.47 to 9.13, respectively.

\begin{table}[htpb]
\centering
\caption{Adaptive-DE advantage over the better local baseline at 30,000 FEs on transfer conditions, reported in spectral-range percentage points. Positive values favor adaptive DE.}
\label{tab:transfer}
\maintablestyle
\begin{tabular}{@{}llrr@{}}
\toprule
Model & Variant & $N=10$ & $N=12$\\
\midrule
fresh spin glass & standard $p=3$ & 0.95 & 0.36\\
 & independent $p=6$ & -1.88 & -1.27\\
 & tied reuse $r=6$ & 5.52 & 2.51\\
 & 2+3-local & 3.32 & 2.82\\
\addlinespace
MaxCut QAOA & standard $p=3$ & 0.00 & 0.00\\
 & independent $p=6$ & -0.97 & -0.40\\
 & tied reuse $r=6$ & 1.68 & 2.77\\
\addlinespace
TFIM VQE & HEA $L=1$ & -0.03 & --\\
 & HEA $L=2$ & -0.33 & --\\
Heisenberg VQE & HEA $L=1$ & 0.00 & --\\
 & HEA $L=2$ & -0.13 & --\\
\bottomrule
\end{tabular}
\end{table}

\begin{figure}[htpb]
\centering
\paperfig{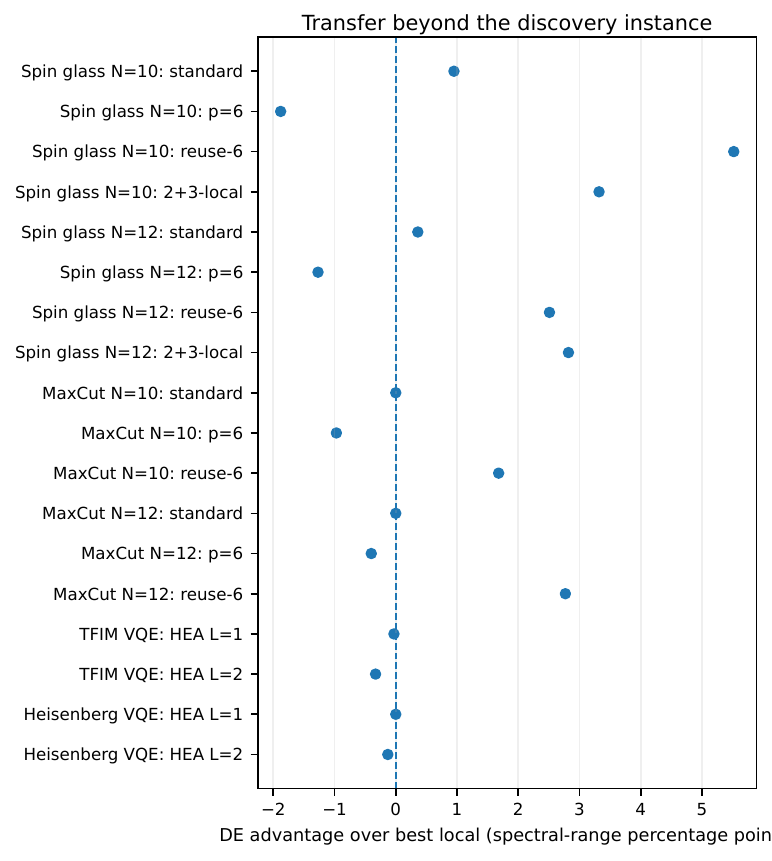}{0.98\linewidth}
\caption{Transfer beyond the discovery instance. Positive values indicate that the descriptively best adaptive-DE method has lower median error than the better local baseline at 30,000 FEs. The standard TFIM and Heisenberg VQE models remain locally accessible, whereas tied parameter reuse shifts both spin-glass and MaxCut QAOA toward global-search advantage.}
\label{fig:transfer}
\end{figure}

The TFIM and Heisenberg VQE controls provide the desired counterexample. Their random-start local searches usually reach endpoints of similar quality: median endpoint gaps range from only 0.12\% to 0.97\% of the spectral range and near-best local-search rates from 53\% to 84\%. Correspondingly, adaptive DE does not improve on MS-BFGS or MS-Powell at 30,000 FEs. The proposed landscape measurements therefore do not label standard physical VQE models as globally difficult merely because they are quantum, noncommuting, or moderately high-dimensional.

\subsection{Landscape measurements prospectively predict the optimizer regime}

\begin{figure}[htpb]
\centering
\paperfig{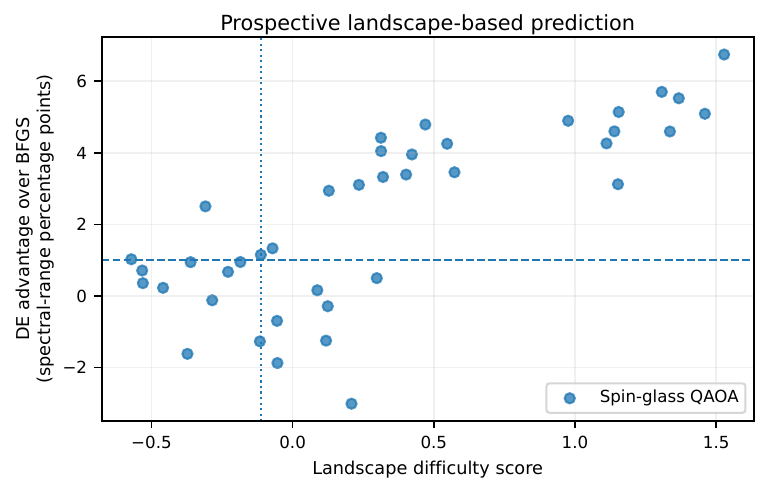}{0.98\linewidth}
\caption{Prospective prediction from landscape diagnostics on 50 new quantum conditions. The score and classification threshold were fixed from Phase II before the Phase-III optimizer outcomes were evaluated. Marker shapes distinguish the four tested model classes. The horizontal dashed line marks the prespecified target of a one-percentage-point spectral-range advantage over MS-BFGS; the vertical dotted line is the Phase-II score threshold.}
\label{fig:prediction}
\end{figure}

The previous results show transfer of specific construction mechanisms. The stronger question is whether the landscape measurements themselves can be used before optimizer benchmarking. Figure~\ref{fig:prediction} compares the precomputed landscape difficulty score with the subsequently observed adaptive-DE advantage at 30,000 FEs.

\begin{figure}[htpb]
\centering
\paperfig{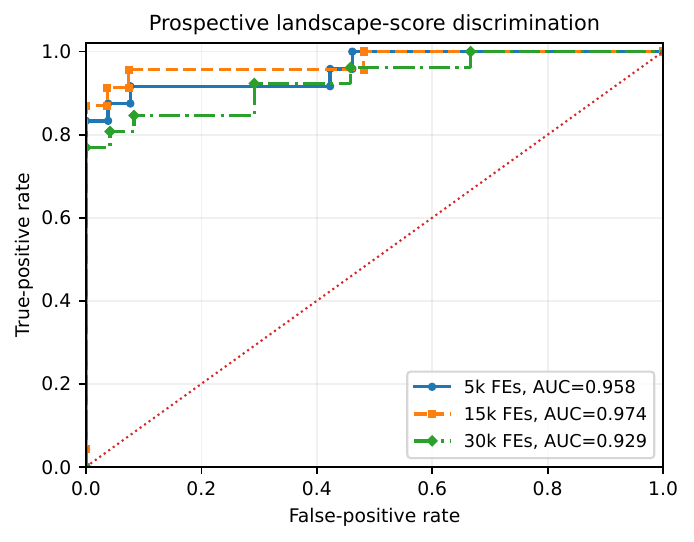}{0.98\linewidth}
\caption{ROC curves for the frozen landscape difficulty score on the 50 Phase-III quantum conditions. AUC measures threshold-independent discrimination of the prespecified target: an adaptive-DE advantage over MS-BFGS larger than one spectral-range percentage point.}
\label{fig:roc}
\end{figure}

Condition-level accuracy is 86\% at 5,000 FEs, 82\% at 15,000 FEs, and 80\% at 30,000 FEs (Table~\ref{tab:prediction}). Balanced accuracy is similar. The continuous score also separates the two classes well without choosing a threshold: ROC AUC is 0.958, 0.974, and 0.929 at the three checkpoints (Figure~\ref{fig:roc}). At 30,000 FEs, a grouped bootstrap over the 14 base instances gives 95\% intervals of $[0.69,0.91]$ for accuracy and $[0.854,0.988]$ for AUC. The score also tracks the magnitude of the optimizer difference: at 30,000 FEs its descriptive Spearman correlation with DE-versus-BFGS advantage is $\rho=0.771$, and with DE advantage over the better of BFGS and Powell it is $\rho=0.758$. Because multiple conditions share the same base instance, we do not attach condition-level independence-based $p$-values to these correlations.

The post-hoc wrapping sensitivity check gives the same qualitative result. Recalibrating on the 12 Phase-II conditions remaining after removal of the incommensurate-mixer and Gaussian-coupling controls produces scores with Spearman correlation $\rho=0.997$ with the frozen primary score at all three checkpoints. Prediction agreement with the primary analysis is 98\%, 92\%, and 92\% at 5,000, 15,000, and 30,000 FEs, respectively. The corresponding accuracies are 88\%, 90\%, and 84\%, with ROC AUC values 0.963, 0.979, and 0.941. Thus the prospective conclusion is insensitive to these two heterogeneous-generator controls. The original 14-condition frozen analysis remains the prespecified primary result.

\begin{table*}[htpb]
\centering
\caption{Condition-level prospective prediction of a DE advantage larger than one spectral-range percentage point on 50 new quantum conditions. AUC is computed from the continuous landscape score before thresholding.}
\label{tab:prediction}
\maintablestyle
\begin{tabular}{@{}rrrrrrr@{}}
\toprule
FE & Accuracy & Balanced acc. & Precision & Recall & Specificity & AUC\\
\midrule
5,000 & 0.860 & 0.862 & 0.815 & 0.917 & 0.808 & 0.958\\
15,000 & 0.820 & 0.830 & 0.733 & 0.957 & 0.704 & 0.974\\
30,000 & 0.800 & 0.796 & 0.767 & 0.885 & 0.708 & 0.929\\
\bottomrule
\end{tabular}
\end{table*}

The errors are themselves informative. At 30,000 FEs all seven false positives are independent-depth $p=6$ QAOA conditions, whereas the three false negatives are standard $p=3$ holdout conditions with modest positive DE advantages. The failure mode is therefore structured rather than random: added oscillatory complexity in deeper QAOA can resemble a globally difficult landscape even when the local-search endpoints remain comparatively similar in quality. This reinforces the controlled-experiment conclusion that multimodality must be interpreted together with basin quality and accessibility.

\section{Discussion}

The combined study answers a more specific question than a conventional optimizer benchmark. Phase II shows that a shift from local-search advantage to global-search advantage can be produced reproducibly by controlled changes to the Hamiltonian--ansatz pair. Phase III then shows that these changes transfer to new instances and that the associated landscape measurements have predictive value outside the discovery panel. The two parts are therefore complementary: construction provides mechanism, while prospective validation tests usefulness.

The strongest mechanism is repeated use of a small parameter set. Tied QAOA repetition increases quantum evolution while keeping the classical search dimension fixed. Repeated occurrences of parameter-dependent generators can enlarge accessible frequency content in related parameterized-circuit settings \cite{Schuld2021}, which is consistent with the observed increase in Fourier complexity. The present results show, however, that frequency growth alone is not enough. Independent $p=6$ depth also increases line complexity but usually remains favorable to local search. The decisive change is the combination of multimodality with a broad distribution of local-search endpoint quality.

This distinction resolves an ambiguity in statements such as ``the landscape has many local minima.'' A landscape can contain many minima and still be easy for multistart local search if a large fraction of random starts reaches good basins or if most minima have similar objective values. What matters operationally is how often local descent reaches a good basin and how costly a bad basin is when it does not. The near-best local-search rate, median endpoint gap, and poor-endpoint fraction directly measure those quantities. Their interpretation is simpler than local curvature metrics and, in this study, more informative for optimizer choice.

Mixed 2-local/3-local costs provide a second mechanism. Their non-monotone dependence on relative scale suggests that competition between structures, rather than interaction order by itself, creates the difficult geometry. This resembles the logic of hybrid and composition benchmarks in classical black-box optimization \cite{awad2017,Novak2026CEC}: difficulty arises from combining structures with different geometry, not from maximizing one scalar notion of complexity.

The eight-instance experiment substantially strengthens the mechanism claim. The same physical Hamiltonian is compared under standard depth, independent depth, tied reuse, and mixed locality, so instance-to-instance disorder is paired out of the comparison. Both difficult constructions improve the global-search advantage on every holdout instance, whereas independent depth reduces it on every instance. This pattern also survives the addition of Powell as a qualitatively different local baseline.

Transfer to MaxCut is particularly useful because it moves the parameter-reuse mechanism away from the frustrated-spin construction in which it was discovered. Standard MaxCut QAOA at the tested sizes is locally accessible, while tied reuse simultaneously increases bad local-search outcomes and makes adaptive DE preferable. Conversely, the TFIM and Heisenberg VQE controls remain favorable to local optimization. The result is therefore not ``global optimizers are better for VQAs,'' but rather that identifiable Hamiltonian--ansatz structures can place a VQA into a different optimization regime.

The prospective experiment is the practical counterpart of the mechanism result. The landscape score is intentionally simple and was not refitted to the new data. Its 80--86\% condition-level accuracy and 0.93--0.97 AUC indicate that a one-time pre-benchmark diagnostic scan contains useful information before committing to an optimizer family. The diagnostic evaluations are separate from the optimizer FE budgets, so this result should be read as evidence for decision value rather than as a claim of net FE savings. Its main error mode is also interpretable: independent depth can raise oscillatory complexity without producing poor basins, reinforcing the need to read corrugation together with basin quality.

The classical controls also illustrate why ``global optimizer'' should not be treated as an intrinsic label of a benchmark function. Rastrigin and Schwefel behave as expected against multistart BFGS, but Powell can be highly competitive in the low dimensions used here. The scientifically relevant comparison is therefore between concrete solver families under a fixed FE budget, not an absolute classification of functions as local or global.

The use of FE rather than wall-clock budgets is deliberate. Within a fixed Hamiltonian--ansatz condition, all methods pay for the same exact objective call, while their Python-level bookkeeping and linear-algebra overheads differ. The study therefore asks which search strategy extracts the best solution from a fixed number of expensive objective evaluations. This is the conventional black-box perspective and is especially relevant to VQAs, where circuit evaluations rather than classical optimizer arithmetic dominate on hardware. Wall-clock efficiency on a particular simulator or device is a separate implementation question.

The present study isolates intrinsic exact-objective geometry on systems up to $N=12$. Finite-shot noise and larger systems add separate sources of difficulty and are natural directions for validation. The prospective score is calibrated from a compact Phase-II panel, so the present claim is correspondingly practical rather than universal: within the tested families, it identifies the optimizer regime well enough to motivate landscape-guided selection. Family-level \emph{best adaptive-DE} summaries remain descriptive by construction; formal pairwise inference is based on named optimizers as specified in Methods.

\paragraph{Future work: VQE and circuit--optimization trade-offs} The present mechanisms are demonstrated most directly for QAOA. A natural next test is a small Hamiltonian-inspired VQE, such as H$_2$ VHA, where parameter sharing can be introduced across repeated noncommuting generator blocks and compared with independently parameterized depth. This would test whether the reuse mechanism is specific to alternating-operator QAOA or reflects a broader property of variational parameterizations. More broadly, the results motivate an ansatz-design question: can some quantum-circuit complexity be traded for a harder classical landscape when strong global optimizers are available? Such a trade-off could be useful if a shallower or more compressed circuit reduces hardware error while preserving sufficient reachability, even at the cost of a less local-search-friendly objective.

\section{Conclusion}

This study uses a landscape-first workflow to ask when global evolutionary search is useful for variational quantum algorithms. Controlled confirmation identifies two mechanisms that make local optimization unreliable: repeated use of a small QAOA parameter set and competition between 2-local and 3-local cost structures. Simply increasing circuit depth and parameter count does not reproduce the effect.

The mechanisms generalize. On eight unseen spin-glass Hamiltonians, both difficult constructions favor adaptive DE over two multistart local baselines on every instance, while independent QAOA depth does not. Tied parameter reuse also transfers to MaxCut at $N=10$ and $N=12$, whereas standard TFIM and Heisenberg VQE models remain locally accessible.

The practical result is that optimizer choice can be approached as a landscape-diagnosis problem rather than by exhaustive benchmarking of unrelated VQAs. Random-start local minimizations reveal whether good basins are easy to reach and whether local-search endpoints differ substantially in quality; random one-dimensional slices provide complementary information about global multimodality. A score formed from these measurements, fixed before the new optimizer runs, predicts whether the best tested adaptive-DE median has a substantive advantage over MS-BFGS on 50 new quantum conditions with 80--86\% condition-level accuracy. Within the tested families, the clearest signal that global search is useful is not circuit depth or local curvature, but a landscape in which local search frequently reaches meaningfully inferior basins.

\section*{Acknowledgments}
This project has received funding from the Research Council of Lithuania (LMTLT), agreement No. P-ITP-24-9. This research was also supported by research grant SGS No. SP2026/063 of VSB--Technical University of Ostrava, Czech Republic.

\section*{Data availability}
The source code required to reproduce the numerical experiments is available at
\url{https://github.com/VojtechNovak/VQA-evolutionary-global}.

\section*{Competing interests}
The authors declare no known competing financial interests or personal relationships that could have influenced the work reported here.

\appendix

\section{Geometric interpretation of the landscape measurements}
\label{app:landscape_geometry}

\begin{figure}[htpb]
\centering
\paperfig{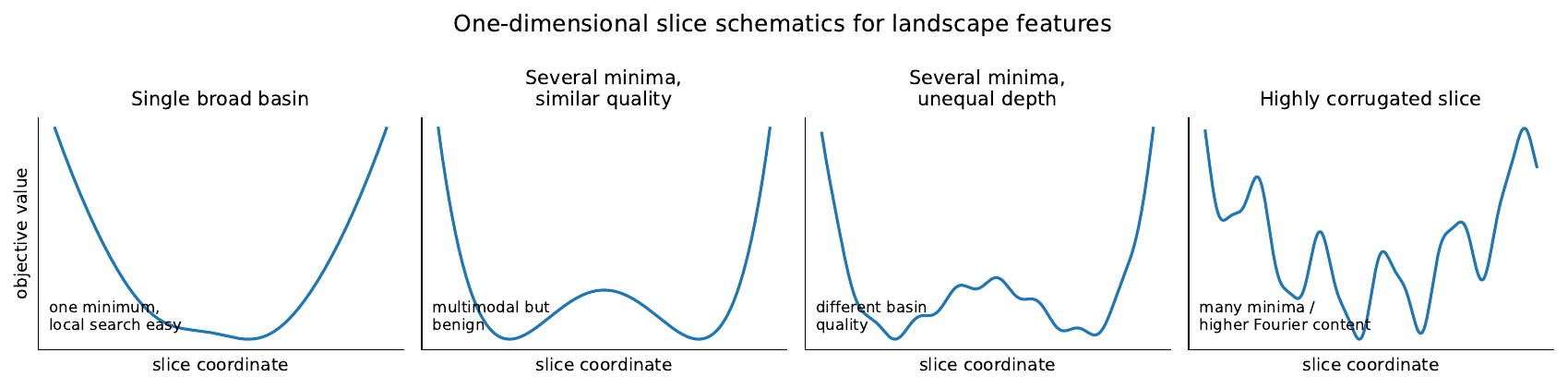}{0.98\linewidth}
\caption{Schematic one-dimensional slices through an objective landscape. A single broad basin is favorable to local search. Multiple minima are not necessarily difficult if their objective values are similar. Difficulty for local optimization increases when different minima have substantially different quality, while a highly corrugated slice raises the counted number of minima and spreads power over more Fourier frequencies. The main-text random-line diagnostics distinguish these cases statistically over many randomly oriented slices.}
\label{fig:diag_slices}
\end{figure}

\begin{figure}[htpb]
\centering
\paperfig{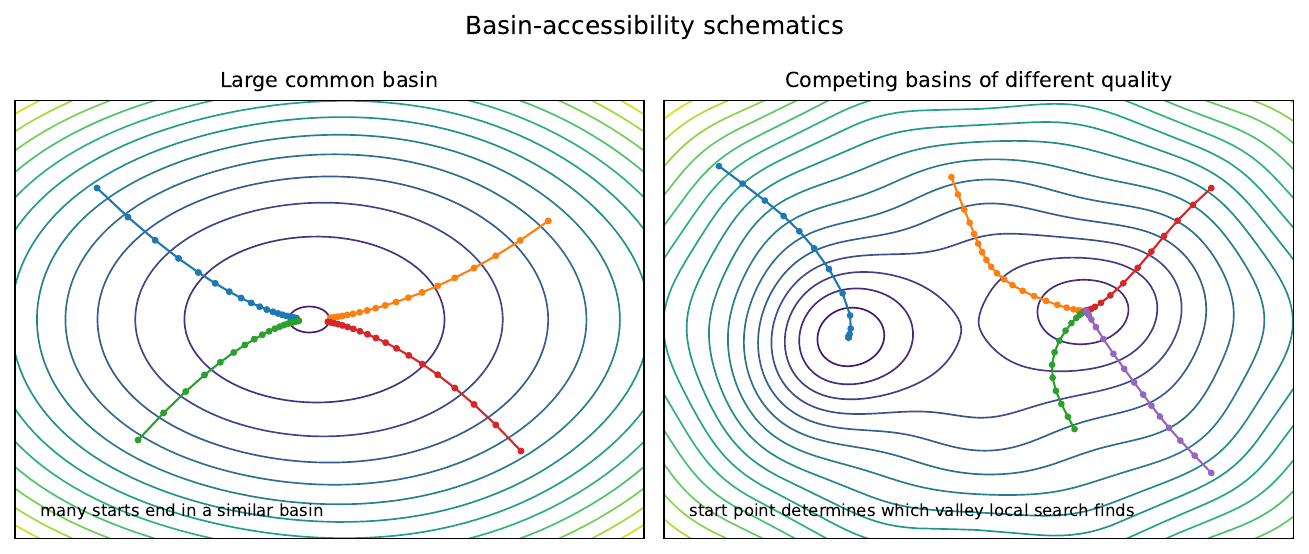}{0.90\linewidth}
\caption{Schematic basin accessibility in two dimensions. In a common-funnel landscape (left), many random starting points descend toward solutions of similar quality, so multistart local search has a high probability of finding a good basin. With competing basins of unequal depth (right), the endpoint depends strongly on the starting point. The near-best local-search rate estimates how often random starts reach good basins, while the endpoint-gap and poor-endpoint measurements quantify differences among the basin minima.}
\label{fig:diag_basins}
\end{figure}

\begin{figure}[htpb]
\centering
\paperfig{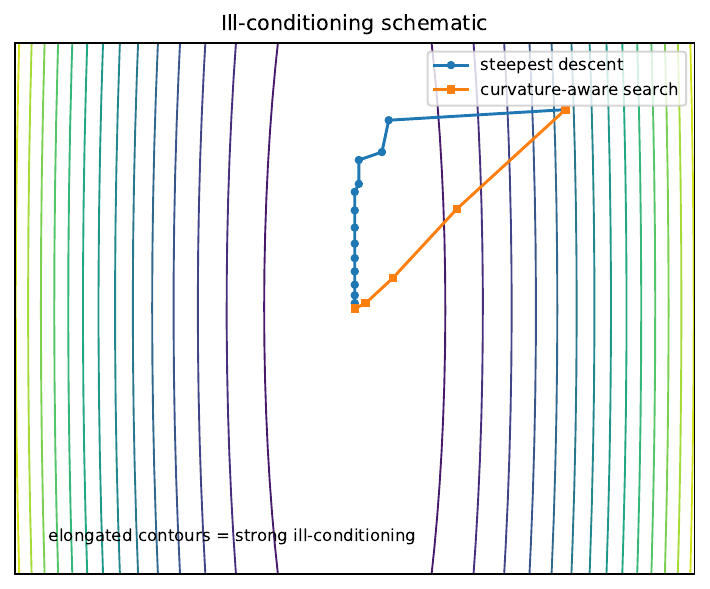}{0.62\linewidth}
\caption{Schematic local ill-conditioning. Strongly elongated contours correspond to very different curvatures in different directions. A simple steepest-descent path can make slow, alternating progress across the narrow valley, while a curvature-aware method can align its steps more effectively. This is a local property: a large local Hessian anisotropy measure says little about whether distant basins have different objective values or how often random starts reach a good basin.}
\label{fig:diag_conditioning}
\end{figure}

The numerical diagnostics in the main text compress high-dimensional parameter-space geometry into scalar summaries. After the main Results show representative real quantum slices, Figures~\ref{fig:diag_slices}--\ref{fig:diag_conditioning} give deliberately simple one- and two-dimensional sketches of what those summaries mean. They are conceptual illustrations, not additional experimental data. Their purpose is to make clear which geometric information a diagnostic captures and, equally importantly, what it does not capture.

Figure~\ref{fig:diag_slices} separates two notions that can otherwise be conflated. The number of minima encountered along a slice measures \emph{corrugation}: how often the surface changes direction along a typical line. The local-search endpoint gap and poor-endpoint fraction instead measure \emph{basin-quality heterogeneity}: whether the valleys reached by local optimization end at comparable or substantially different objective values. A rugged surface with many nearly equivalent minima can therefore be easier than a less oscillatory surface containing a few broad but inferior basins.

The basin view in Figure~\ref{fig:diag_basins} explains why we repeatedly minimize from random starting points rather than merely count stationary points. For finite-budget optimization, a minimum matters in proportion to both its quality and how often the specified random-start local search reaches it. An excellent basin that is rarely reached can be practically inaccessible under a finite budget, whereas a good basin reached from many starts can make a formally multimodal landscape easy.

Figure~\ref{fig:diag_conditioning} also clarifies why the positive Hessian eigenvalue ratio is a weaker predictor in our experiments. This ratio summarizes positive-curvature anisotropy near one endpoint and can affect the path taken inside one valley, but it does not describe the number, relative depths, or accessibility of other valleys. The independent-depth QAOA control provides the empirical example: it can have strongly anisotropic local curvature and a corrugated landscape while remaining comparatively favorable to multistart local search.

\section{Optimizer definitions and implementation details}
\label{app:optimizers}

This appendix gives the algorithmic and implementation detail behind the compact rationale in Table~\ref{tab:optimizer_rationale}. The purpose is reproducibility and interpretation, not to claim that any one implementation is universally representative of its optimizer family. All methods minimize exactly the same objective under the same FE ceiling. Every call to the objective is counted, including evaluations used internally by numerical gradients, line searches, or population initialization.

\begin{figure*}[htpb]
\centering
\paperfig{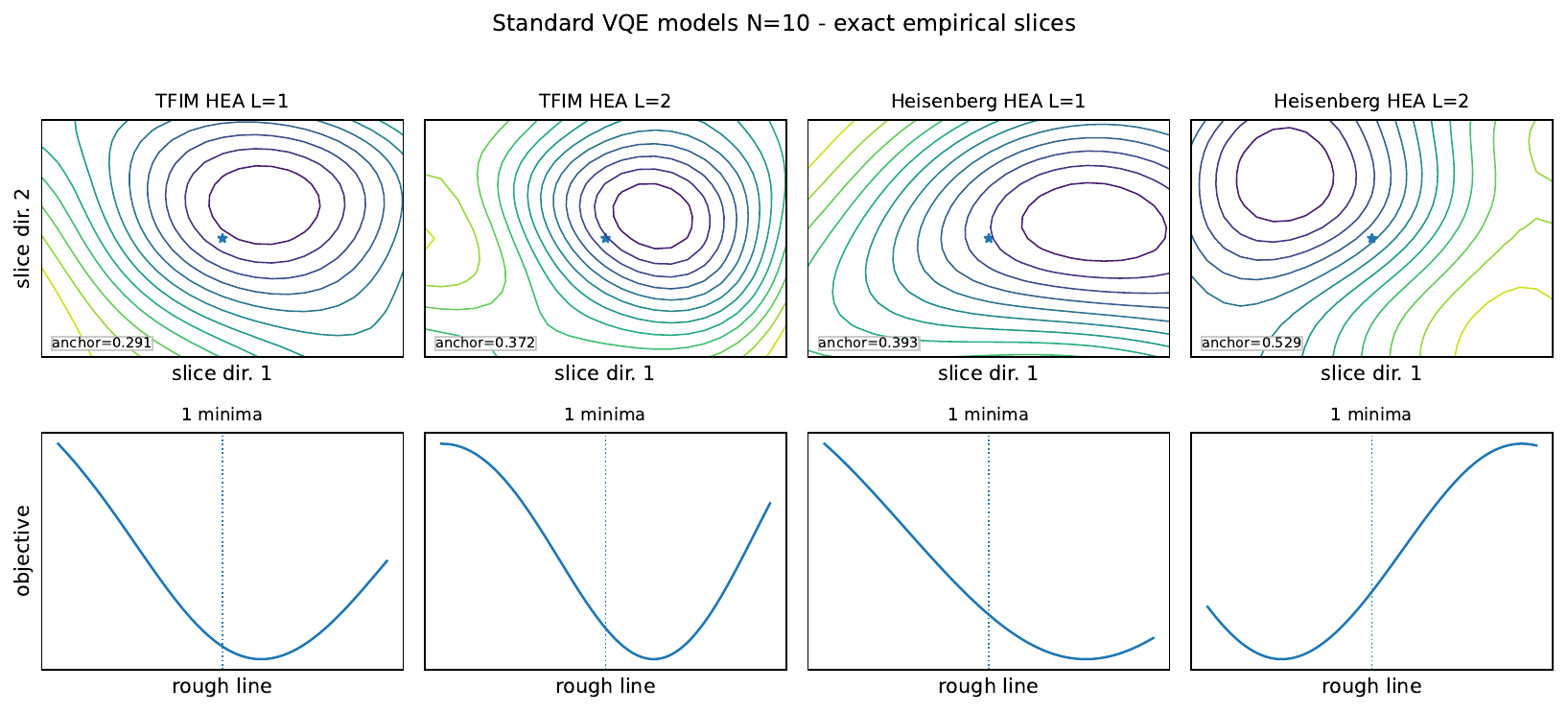}{0.65\textwidth}
\caption{High-corrugation scans for the standard VQE controls: TFIM and Heisenberg Hamiltonians with hardware-efficient ansatzes at $L=1,2$. Even after deliberately selecting the roughest of six random directions, all four one-dimensional slices contain a single minimum, consistent with their local-search-friendly optimizer behavior.}
\label{fig:empirical_vqe}
\end{figure*}

\subsection{Shared search domain, seeding, and budget accounting}

All optimizer coordinates are represented in the common bounded box $[-\pi,\pi)^D$, and random initial points are sampled uniformly from it. Population methods are supplied these numerical bounds, while evaluations proposed outside the box by local methods are mapped back coordinatewise by the common wrapper. This is an explicit benchmark boundary convention; for heterogeneous generator weights it is not interpreted as a statement of exact $2\pi$ physical periodicity.

Replicate index and integer seed labels are shared across optimizers within a condition. Exact statevector objectives are deterministic; the seed controls stochastic optimizer choices and random initial points, but different algorithms consume random numbers differently, so this blocking does not imply a common random-number path. Population sizes are kept moderate relative to the FE budget. The wrapper applies an explicit population cap of 150 when the optimizer interface exposes a writable population-size argument. For externally supplied implementations whose interface does not expose such an argument, the optimizer's native population rule is retained. In all cases, the common objective wrapper enforces the FE ceiling exactly.

\subsection{Multistart BFGS}

BFGS is a quasi-Newton local method. Starting from $x_k$, it builds an approximation to local curvature from successive steps and gradient estimates, and uses this information to choose a descent direction \cite{NocedalWright2006}. Because a single BFGS trajectory explores only one basin of attraction, the benchmark repeatedly restarts BFGS from independent uniformly sampled points until the common FE budget is exhausted. This makes MS-BFGS a direct test of whether repeated local optimization can reliably locate a good basin.

We use \texttt{scipy.optimize.minimize(method="BFGS")} with a maximum of 1,800 iterations per restart and gradient tolerance $10^{-8}$. No analytic Jacobian and no custom finite-difference step are supplied, so SciPy's numerical-gradient rule is used and all objective calls generated by it count toward the FE budget. This is distinct from the finite-difference steps used separately for landscape diagnostics.

\subsection{Multistart Powell}

Powell's method is a derivative-free local direction-set algorithm. It performs sequential one-dimensional searches along a set of directions and updates those directions using the displacement achieved over a cycle \cite{Powell1964}. It therefore provides a qualitatively different local baseline from BFGS and does not depend on finite-difference gradients. The Phase-III wrapper launches independent random starts until the common FE budget is exhausted. We use \texttt{scipy.optimize.minimize(method="Powell")} with a maximum of 1,800 iterations per restart, parameter tolerance $10^{-8}$, and function tolerance $10^{-10}$. MS-Powell was added only in Phase III, so it is used as an out-of-sample robustness baseline rather than retroactively entering mechanism selection.

\subsection{Multistart CMA-ES}

CMA-ES samples a population from a multivariate normal search distribution,
$x\sim\mathcal{N}(m,\sigma^2 C)$, and adapts the mean, covariance matrix, and global step size from successful samples \cite{HansenOstermeier2001}. Covariance adaptation allows the population to rotate and stretch along favorable directions without derivative information. It is included as a population-based comparison with a mechanism substantially different from DE.

Our multistart wrapper uses initial step size $\sigma_0=1$, bounds $[-\pi,\pi]^D$, parameter and objective tolerances $10^{-10}$, and population size
\begin{equation}
\lambda=\min\!\left(150,\;4+\left\lfloor 3\ln(\max(D,2))\right\rfloor\right).
\end{equation}
For the main $D=6$ QAOA conditions this gives $\lambda=9$. After a CMA-ES trajectory terminates, the method is restarted from a new uniformly sampled point if FE budget remains.

\subsection{Adaptive Differential Evolution family}

Differential Evolution (DE) evolves a population by constructing donor vectors from scaled differences between population members, recombining donors with target vectors, and greedily retaining improved trial points \cite{StornPrice1997}. Modern adaptive variants modify the mutation scale $F$, crossover rate $CR$, population size, and selection pressure during the run. We use three such variants so that the main conclusion does not depend on a single DE implementation.

SHADE introduced success-history memories that bias future $F$ and $CR$ values toward settings that recently generated improvements \cite{TanabeFukunaga2013}; L-SHADE added linear population-size reduction \cite{TanabeFukunaga2014}. The three methods below belong to this general lineage but use different adaptation rules.

\paragraph{iL-SHADE.}
iL-SHADE is an improved L-SHADE variant for real-parameter optimization \cite{Brest2016iLSHADE}. We use the \texttt{pyade.ilshade} implementation with its standard adaptive mechanisms and explicitly set
\[
NP_{\rm init}=\min(150,\max(8,4D)).
\]
Thus the main $D=6$ conditions start from $NP_{\rm init}=24$. The implementation uses success-history parameter adaptation, an external archive, and linear population reduction; the remaining internal settings are the library defaults used by the experiment code. One adaptive trajectory runs until the FE budget is exhausted.

\paragraph{jSO (streamlined implementation).}
The supplied \texttt{jso} module is a streamlined variant of the CEC-2017 jSO family \cite{Brest2017jSO}. We use it simply as a second strong adaptive-DE reference, not to study the effect of individual jSO design choices. It retains the characteristic current-to-$p$best-style search, adaptive parameter memories, and population reduction, while omitting some of the stage-dependent restrictions of canonical jSO.

Its documented native initialization rule is
\begin{equation}
NP_{\rm init}
=
\max\!\left(30,\left\lfloor25\sqrt{D}\log_{10}D\right\rfloor\right).
\end{equation}
For $D=6$, $25\sqrt{6}\log_{10}6=47.65\ldots$, so this rule gives $NP_{\rm init}=47$ (not 30). The generic wrapper also records a requested population target for consistency with the other population methods; in the completed runs the supplied jSO interface did not expose a separate writable population-size control, so no external override was recorded and the module's native setting was used. Figures and result tables keep the short label ``jSO'' to match the optimizer name stored by the code.

\paragraph{L-SRTDE.}
L-SRTDE is a recent adaptive DE variant whose mutation scaling responds to the current success rate, i.e. the fraction of trial vectors that improve their parents, together with population-size reduction and adaptive crossover control \cite{Stanovov2024LSRTDE}. In the supplied CEC-style implementation, the native initial population scales as $20D$ and is reduced toward a small terminal population. For the main $D=6$ conditions this corresponds to an initial target of 120 individuals. The wrapper requests
\[
NP_{\rm init}=\min(150,\max(20,20D))
\]
when a writable population-size argument is exposed; otherwise the implementation's native rule is retained. One adaptive trajectory runs until the FE budget is exhausted.

\begin{table*}[h]
\centering
\caption{Optimizer implementation settings used in the study. ``Restart'' means that a fresh independent run is launched while FE budget remains. Population rules listed as native are controlled inside the supplied optimizer module rather than by the common wrapper.}
\label{tab:optimizer_settings}
\footnotesize
\begin{tabular}{p{2.4cm}p{4.5cm}p{6.0cm}p{2.4cm}}
\toprule
Method & Population / initialization & Main settings used here & Budget behavior / implementation \\
\midrule
MS-BFGS &
$x_0\sim\mathcal{U}[-\pi,\pi)^D$ &
SciPy BFGS; no analytic Jacobian; max. 1,800 iterations per local run; $g_{\rm tol}=10^{-8}$; numerical-gradient and line-search calls count as FEs. &
Restart to FE limit; SciPy. \\

MS-Powell &
$x_0\sim\mathcal{U}[-\pi,\pi)^D$ &
SciPy Powell; max. 1,800 iterations; $x_{\rm tol}=10^{-8}$; $f_{\rm tol}=10^{-10}$. &
Restart to FE limit; SciPy; Phase III only. \\

MS-CMA-ES &
$\lambda=\min(150,4+\lfloor3\ln(\max(D,2))\rfloor)$; $\lambda=9$ at $D=6$ &
$\sigma_0=1$; bounded box; parameter and objective tolerances $10^{-10}$. &
Restart to FE limit; supplied \texttt{CMAES.py}. \\

iL-SHADE &
$NP_{\rm init}=\min(150,\max(8,4D))$; 24 at $D=6$ &
Success-history adaptation, external archive, and population reduction; remaining settings from \texttt{pyade.ilshade} defaults. &
One adaptive trajectory; PyADE. \\

jSO &
Native $NP_{\rm init}=\max(30,\lfloor25\sqrt D\log_{10}D\rfloor)$; 47 at $D=6$ &
Streamlined jSO implementation; current-to-$p$best-style search, adaptive $F/CR$ memories, population reduction; native internal control retained. &
One adaptive trajectory; supplied jSO module. \\

L-SRTDE &
Native scaling $NP_{\rm init}\approx20D$; 120 at $D=6$ &
Success-rate-dependent mutation scaling, adaptive crossover control, and population reduction; wrapper requests a cap when exposed. &
One adaptive trajectory; supplied L-SRTDE module. \\
\bottomrule
\end{tabular}
\end{table*}

The three DE variants are not treated as three independent scientific replications. They are alternative algorithms within one global-search family. This is why $\DEadvB$ and $\DEadvL$ are explicitly descriptive family summaries, whereas formal paired tests in the manuscript compare named algorithms with the local baseline.

\section{Complete Phase-II confirmation summary}

Table~\ref{tab:all_phase2} reports every Phase-II condition through the three descriptive adaptive-DE advantages in the underlying normalized-error units. This appendix is intended as a complete condition-level audit of the controlled confirmation experiment; the main text converts the same differences to spectral-range percentage points for readability. Positive values favor the best adaptive-DE median over MS-BFGS.

\begin{table*}[t]
\centering
\caption{Complete Phase-II confirmation results.}
\label{tab:all_phase2}
\scriptsize
\setlength{\tabcolsep}{5pt}
\begin{tabular}{@{}lrrrrl@{}}
\toprule
Condition & $D$ & $\DEadvB(5k)$ & $\DEadvB(15k)$ & $\DEadvB(30k)$ & Best DE at 30k\\
\midrule
Ising spin-glass reference, $N=10$ & 10 & -0.009 & -0.002 & -0.002 & jSO \\
Ising spin-glass reference, $N=12$ & 12 & -6.4e-09 & -0.001 & 7.2e-16 & jSO \\
hardware-efficient ansatz, reuse 3 & 20 & -0.189 & -0.038 & -0.010 & L-SRTDE \\
QAOA $p=3$ & 6 & 0.003 & 0.005 & 0.005 & jSO \\
QAOA $p=6$ & 12 & -0.038 & -0.018 & -4.4e-04 & L-SRTDE \\
QAOA $p=3$, reuse 2 & 6 & 0.024 & 0.020 & 0.018 & jSO \\
QAOA $p=3$, reuse 4 & 6 & 0.042 & 0.041 & 0.018 & L-SRTDE \\
QAOA $p=3$, reuse 6 & 6 & 0.077 & 0.051 & 0.040 & L-SRTDE \\
2+3-local, $\alpha_3=0.5$ & 6 & 0.017 & 0.033 & 0.029 & L-SRTDE \\
2+3-local, $\alpha_3=1$ & 6 & 0.041 & 0.052 & 0.031 & L-SRTDE \\
2+3-local, $\alpha_3=2$ & 6 & 0.025 & 0.039 & 0.027 & L-SRTDE \\
incommensurate mixer & 6 & 0.006 & 0.024 & 0.015 & jSO \\
binary couplings & 6 & -0.016 & -0.006 & -6.3e-05 & jSO \\
Gaussian couplings & 6 & 3.4e-04 & 0.001 & 0.017 & jSO \\
Rastrigin $D=6$ & 6 & 0.054 & 0.050 & 0.025 & L-SRTDE \\
Rastrigin $D=10$ & 10 & 0.056 & 0.089 & 0.080 & jSO \\
Schwefel $D=6$ & 6 & 0.069 & 0.056 & 0.039 & L-SRTDE \\
Schwefel $D=10$ & 10 & 0.007 & 0.067 & 0.063 & jSO \\
\bottomrule
\end{tabular}
\end{table*}

The full table shows that the main-text pattern is not created by omitting inconvenient controls. All tied-reuse and all nonzero mixed-locality conditions have positive DE advantages at every checkpoint. In contrast, the independent-depth and hardware-efficient controls are non-positive at 30,000 FEs, and the coupling/mixer variants are smaller or less systematic. The classical controls verify that the optimizer implementations can exhibit a conventional multimodal local/global reversal, but they are not used to define the quantum mechanisms.

\section{Selected Phase-II paired optimizer comparisons}

Positive paired advantage means lower error for the listed alternative. $p_{\rm Holm}$ is the paired Wilcoxon $p$-value after Holm correction within a condition and checkpoint.

\begin{table*}[htpb]
\centering
\scriptsize
\begin{tabular}{lrlrrrr}
\toprule
Condition & FE & Alternative & Paired advantage & 95\% CI & Win frac. & $p_{\rm Holm}$\\
\midrule
QAOA $p=3$, reuse 2 & 5,000 & iL-SHADE & 0.023 & [0.010, 0.046] & 0.93 & 2.44e-04 \\
QAOA $p=3$, reuse 2 & 30,000 & jSO & 0.015 & [0.010, 0.022] & 0.93 & 4.88e-04 \\
QAOA $p=3$, reuse 4 & 5,000 & iL-SHADE & 0.040 & [0.011, 0.058] & 1.00 & 2.44e-04 \\
QAOA $p=3$, reuse 4 & 30,000 & jSO & 0.015 & [0.007, 0.031] & 0.93 & 4.88e-04 \\
QAOA $p=3$, reuse 6 & 5,000 & iL-SHADE & 0.077 & [0.066, 0.088] & 1.00 & 2.44e-04 \\
QAOA $p=3$, reuse 6 & 30,000 & L-SRTDE & 0.038 & [0.034, 0.042] & 1.00 & 2.44e-04 \\
2+3-local, $\alpha_3=0.5$ & 5,000 & iL-SHADE & 0.029 & [0.019, 0.046] & 0.87 & 0.003 \\
2+3-local, $\alpha_3=0.5$ & 30,000 & L-SRTDE & 0.024 & [0.021, 0.035] & 1.00 & 2.44e-04 \\
2+3-local, $\alpha_3=1$ & 5,000 & iL-SHADE & 0.047 & [0.027, 0.052] & 0.93 & 4.88e-04 \\
2+3-local, $\alpha_3=1$ & 30,000 & L-SRTDE & 0.027 & [0.026, 0.036] & 1.00 & 2.44e-04 \\
2+3-local, $\alpha_3=2$ & 5,000 & iL-SHADE & 0.029 & [0.013, 0.044] & 0.93 & 4.88e-04 \\
2+3-local, $\alpha_3=2$ & 30,000 & L-SRTDE & 0.028 & [0.024, 0.035] & 1.00 & 2.44e-04 \\
\bottomrule
\end{tabular}%
\end{table*}

These named-optimizer comparisons provide the inferential counterpart to the descriptive best-DE summaries. The largest tied-reuse and mixed-locality effects are accompanied by high paired win fractions, and the same optimizer-run pairing is used at every checkpoint. The table is intentionally selective in print to keep the appendix readable; the complete paired-test output for every condition, checkpoint, and local baseline is retained in the machine-readable results described in the reproducibility appendix.

\section{Complete Phase-III final-condition summary}

Table~\ref{tab:all_phase3} reports the final 30,000-FE family-level comparisons for all 54 retained Phase-III conditions in normalized-error units. It complements the aggregated holdout and transfer results in the main text by showing every individual condition, including the four classical sanity controls. Complete named-optimizer medians, paired tests, and run-level values are retained in the generated machine-readable tables.

\begin{table*}[t]
\centering
\caption{Complete Phase-III results at 30,000 FEs.}
\label{tab:all_phase3}
\tiny
\renewcommand{\arraystretch}{0.93}
\setlength{\tabcolsep}{3.5pt}
\begin{tabular}{@{}lllrrr@{}}
\toprule
Model & Instance & Variant & $D$ & $\DEadvB$ & $\DEadvL$\\
\midrule
holdout spin glass & H01 & 2+3-local & 6 & 0.040 & 0.040 \\
holdout spin glass & H01 & independent $p=6$ & 12 & -0.007 & -0.007 \\
holdout spin glass & H01 & standard $p=3$ & 6 & 0.010 & 0.010 \\
holdout spin glass & H01 & tied reuse $r=6$ & 6 & 0.043 & 0.043 \\
holdout spin glass & H02 & 2+3-local & 6 & 0.040 & 0.040 \\
holdout spin glass & H02 & independent $p=6$ & 12 & -0.001 & -0.001 \\
holdout spin glass & H02 & standard $p=3$ & 6 & 0.007 & 0.007 \\
holdout spin glass & H02 & tied reuse $r=6$ & 6 & 0.049 & 0.043 \\
holdout spin glass & H03 & 2+3-local & 6 & 0.031 & 0.031 \\
holdout spin glass & H03 & independent $p=6$ & 12 & -0.012 & -0.012 \\
holdout spin glass & H03 & standard $p=3$ & 6 & 0.009 & 0.009 \\
holdout spin glass & H03 & tied reuse $r=6$ & 6 & 0.067 & 0.051 \\
holdout spin glass & H04 & 2+3-local & 6 & 0.044 & 0.044 \\
holdout spin glass & H04 & independent $p=6$ & 12 & 0.005 & 0.005 \\
holdout spin glass & H04 & standard $p=3$ & 6 & 0.025 & 0.025 \\
holdout spin glass & H04 & tied reuse $r=6$ & 6 & 0.051 & 0.051 \\
holdout spin glass & H05 & 2+3-local & 6 & 0.035 & 0.035 \\
holdout spin glass & H05 & independent $p=6$ & 12 & 0.002 & 0.002 \\
holdout spin glass & H05 & standard $p=3$ & 6 & 0.013 & 0.013 \\
holdout spin glass & H05 & tied reuse $r=6$ & 6 & 0.046 & 0.046 \\
holdout spin glass & H06 & 2+3-local & 6 & 0.048 & 0.048 \\
holdout spin glass & H06 & independent $p=6$ & 12 & -0.030 & -0.030 \\
holdout spin glass & H06 & standard $p=3$ & 6 & 0.007 & 0.007 \\
holdout spin glass & H06 & tied reuse $r=6$ & 6 & 0.046 & 0.044 \\
holdout spin glass & H07 & 2+3-local & 6 & 0.034 & 0.030 \\
holdout spin glass & H07 & independent $p=6$ & 12 & -0.016 & -0.016 \\
holdout spin glass & H07 & standard $p=3$ & 6 & 0.002 & 0.002 \\
holdout spin glass & H07 & tied reuse $r=6$ & 6 & 0.051 & 0.051 \\
holdout spin glass & H08 & 2+3-local & 6 & 0.043 & 0.043 \\
holdout spin glass & H08 & independent $p=6$ & 12 & -0.003 & -0.003 \\
holdout spin glass & H08 & standard $p=3$ & 6 & 0.011 & 0.011 \\
holdout spin glass & H08 & tied reuse $r=6$ & 6 & 0.057 & 0.033 \\
size-transfer spin glass & N=10 & 2+3-local & 6 & 0.033 & 0.033 \\
size-transfer spin glass & N=10 & independent $p=6$ & 12 & -0.019 & -0.019 \\
size-transfer spin glass & N=10 & standard $p=3$ & 6 & 0.009 & 0.009 \\
size-transfer spin glass & N=10 & tied reuse $r=6$ & 6 & 0.055 & 0.055 \\
size-transfer spin glass & N=12 & 2+3-local & 6 & 0.029 & 0.028 \\
size-transfer spin glass & N=12 & independent $p=6$ & 12 & -0.013 & -0.013 \\
size-transfer spin glass & N=12 & standard $p=3$ & 6 & 0.004 & 0.004 \\
size-transfer spin glass & N=12 & tied reuse $r=6$ & 6 & 0.031 & 0.025 \\
Heisenberg VQE & N=10 & HEA $L=1$ & 20 & 0.000 & 0.000 \\
Heisenberg VQE & N=10 & HEA $L=2$ & 30 & -0.001 & -0.001 \\
MaxCut QAOA & N=10 & independent $p=6$ & 12 & -0.010 & -0.010 \\
MaxCut QAOA & N=10 & standard $p=3$ & 6 & 0.000 & 0.000 \\
MaxCut QAOA & N=10 & tied reuse $r=6$ & 6 & 0.023 & 0.017 \\
MaxCut QAOA & N=12 & independent $p=6$ & 12 & -0.004 & -0.004 \\
MaxCut QAOA & N=12 & standard $p=3$ & 6 & 0.000 & 0.000 \\
MaxCut QAOA & N=12 & tied reuse $r=6$ & 6 & 0.029 & 0.028 \\
TFIM VQE & N=10 & HEA $L=1$ & 20 & 0.000 & 0.000 \\
TFIM VQE & N=10 & HEA $L=2$ & 30 & -0.003 & -0.003 \\
Rastrigin & D=10 & control & 10 & 0.080 & 0.009 \\
Rastrigin & D=6 & control & 6 & 0.048 & 0.002 \\
Schwefel & D=10 & control & 10 & 0.069 & 0.033 \\
Schwefel & D=6 & control & 6 & 0.020 & 0.000 \\
\bottomrule
\end{tabular}
\end{table*}

The condition-level results reinforce three main conclusions. First, tied reuse and mixed locality remain positive across every independent holdout spin glass and both retained system sizes. Second, independent $p=6$ depth is consistently neutral or favorable to local search despite its increased line complexity. Third, the transfer controls behave sensibly: tied reuse moves MaxCut into a global-search-favorable regime, whereas the standard TFIM and Heisenberg VQE cases remain near zero or negative. The classical rows additionally show why Powell was useful as a second local baseline: its inclusion can substantially reduce the apparent global-search advantage on low-dimensional Rastrigin and Schwefel.

\clearpage
\bibliography{references}

\begin{thebibliography}{42}%
\makeatletter
\providecommand \@ifxundefined [1]{%
 \@ifx{#1\undefined}
}%
\providecommand \@ifnum [1]{%
 \ifnum #1\expandafter \@firstoftwo
 \else \expandafter \@secondoftwo
 \fi
}%
\providecommand \@ifx [1]{%
 \ifx #1\expandafter \@firstoftwo
 \else \expandafter \@secondoftwo
 \fi
}%
\providecommand \natexlab [1]{#1}%
\providecommand \enquote  [1]{``#1''}%
\providecommand \bibnamefont  [1]{#1}%
\providecommand \bibfnamefont [1]{#1}%
\providecommand \citenamefont [1]{#1}%
\providecommand \href@noop [0]{\@secondoftwo}%
\providecommand \href [0]{\begingroup \@sanitize@url \@href}%
\providecommand \@href[1]{\@@startlink{#1}\@@href}%
\providecommand \@@href[1]{\endgroup#1\@@endlink}%
\providecommand \@sanitize@url [0]{\catcode `\\12\catcode `\$12\catcode `\&12\catcode `\#12\catcode `\^12\catcode `\_12\catcode `\%12\relax}%
\providecommand \@@startlink[1]{}%
\providecommand \@@endlink[0]{}%
\providecommand \url  [0]{\begingroup\@sanitize@url \@url }%
\providecommand \@url [1]{\endgroup\@href {#1}{\urlprefix }}%
\providecommand \urlprefix  [0]{URL }%
\providecommand \Eprint [0]{\href }%
\providecommand \doibase [0]{https://doi.org/}%
\providecommand \selectlanguage [0]{\@gobble}%
\providecommand \bibinfo  [0]{\@secondoftwo}%
\providecommand \bibfield  [0]{\@secondoftwo}%
\providecommand \translation [1]{[#1]}%
\providecommand \BibitemOpen [0]{}%
\providecommand \bibitemStop [0]{}%
\providecommand \bibitemNoStop [0]{.\EOS\space}%
\providecommand \EOS [0]{\spacefactor3000\relax}%
\providecommand \BibitemShut  [1]{\csname bibitem#1\endcsname}%
\let\auto@bib@innerbib\@empty
\bibitem [{\citenamefont {Peruzzo}\ \emph {et~al.}(2014)\citenamefont {Peruzzo}, \citenamefont {McClean}, \citenamefont {Shadbolt}, \citenamefont {Yung}, \citenamefont {Zhou}, \citenamefont {Love}, \citenamefont {Aspuru-Guzik},\ and\ \citenamefont {O'Brien}}]{Peruzzo2014}%
  \BibitemOpen
  \bibfield  {author} {\bibinfo {author} {\bibfnamefont {A.}~\bibnamefont {Peruzzo}}, \bibinfo {author} {\bibfnamefont {J.}~\bibnamefont {McClean}}, \bibinfo {author} {\bibfnamefont {P.}~\bibnamefont {Shadbolt}}, \bibinfo {author} {\bibfnamefont {M.-H.}\ \bibnamefont {Yung}}, \bibinfo {author} {\bibfnamefont {X.-Q.}\ \bibnamefont {Zhou}}, \bibinfo {author} {\bibfnamefont {P.~J.}\ \bibnamefont {Love}}, \bibinfo {author} {\bibfnamefont {A.}~\bibnamefont {Aspuru-Guzik}},\ and\ \bibinfo {author} {\bibfnamefont {J.~L.}\ \bibnamefont {O'Brien}},\ }\bibfield  {title} {\bibinfo {title} {A variational eigenvalue solver on a photonic quantum processor},\ }\href {https://doi.org/10.1038/ncomms5213} {\bibfield  {journal} {\bibinfo  {journal} {Nature Communications}\ }\textbf {\bibinfo {volume} {5}},\ \bibinfo {pages} {4213} (\bibinfo {year} {2014})}\BibitemShut {NoStop}%
\bibitem [{\citenamefont {Farhi}\ \emph {et~al.}(2014)\citenamefont {Farhi}, \citenamefont {Goldstone},\ and\ \citenamefont {Gutmann}}]{Farhi2014}%
  \BibitemOpen
  \bibfield  {author} {\bibinfo {author} {\bibfnamefont {E.}~\bibnamefont {Farhi}}, \bibinfo {author} {\bibfnamefont {J.}~\bibnamefont {Goldstone}},\ and\ \bibinfo {author} {\bibfnamefont {S.}~\bibnamefont {Gutmann}},\ }\bibfield  {title} {\bibinfo {title} {A quantum approximate optimization algorithm},\ }\href@noop {} {\bibfield  {journal} {\bibinfo  {journal} {arXiv preprint arXiv:1411.4028}\ } (\bibinfo {year} {2014})},\ \Eprint {https://arxiv.org/abs/1411.4028} {arXiv:1411.4028 [quant-ph]} \BibitemShut {NoStop}%
\bibitem [{\citenamefont {Cerezo}\ \emph {et~al.}(2021{\natexlab{a}})\citenamefont {Cerezo}, \citenamefont {Arrasmith}, \citenamefont {Babbush}, \citenamefont {Benjamin}, \citenamefont {Endo}, \citenamefont {Fujii}, \citenamefont {McClean}, \citenamefont {Mitarai}, \citenamefont {Yuan}, \citenamefont {Cincio},\ and\ \citenamefont {Coles}}]{Cerezo2021Review}%
  \BibitemOpen
  \bibfield  {author} {\bibinfo {author} {\bibfnamefont {M.}~\bibnamefont {Cerezo}}, \bibinfo {author} {\bibfnamefont {A.}~\bibnamefont {Arrasmith}}, \bibinfo {author} {\bibfnamefont {R.}~\bibnamefont {Babbush}}, \bibinfo {author} {\bibfnamefont {S.~C.}\ \bibnamefont {Benjamin}}, \bibinfo {author} {\bibfnamefont {S.}~\bibnamefont {Endo}}, \bibinfo {author} {\bibfnamefont {K.}~\bibnamefont {Fujii}}, \bibinfo {author} {\bibfnamefont {J.~R.}\ \bibnamefont {McClean}}, \bibinfo {author} {\bibfnamefont {K.}~\bibnamefont {Mitarai}}, \bibinfo {author} {\bibfnamefont {X.}~\bibnamefont {Yuan}}, \bibinfo {author} {\bibfnamefont {L.}~\bibnamefont {Cincio}},\ and\ \bibinfo {author} {\bibfnamefont {P.~J.}\ \bibnamefont {Coles}},\ }\bibfield  {title} {\bibinfo {title} {Variational quantum algorithms},\ }\href {https://doi.org/10.1038/s42254-021-00348-9} {\bibfield  {journal} {\bibinfo  {journal} {Nature Reviews Physics}\ }\textbf {\bibinfo {volume} {3}},\ \bibinfo {pages} {625} (\bibinfo {year}
  {2021}{\natexlab{a}})}\BibitemShut {NoStop}%
\bibitem [{\citenamefont {Tilly}\ \emph {et~al.}(2022)\citenamefont {Tilly}, \citenamefont {Chen}, \citenamefont {Cao}, \citenamefont {Picozzi}, \citenamefont {Setia}, \citenamefont {Li}, \citenamefont {Grant}, \citenamefont {Wossnig}, \citenamefont {Rungger}, \citenamefont {Booth},\ and\ \citenamefont {Tennyson}}]{Tilly2022}%
  \BibitemOpen
  \bibfield  {author} {\bibinfo {author} {\bibfnamefont {J.}~\bibnamefont {Tilly}}, \bibinfo {author} {\bibfnamefont {H.}~\bibnamefont {Chen}}, \bibinfo {author} {\bibfnamefont {S.}~\bibnamefont {Cao}}, \bibinfo {author} {\bibfnamefont {D.}~\bibnamefont {Picozzi}}, \bibinfo {author} {\bibfnamefont {K.}~\bibnamefont {Setia}}, \bibinfo {author} {\bibfnamefont {Y.}~\bibnamefont {Li}}, \bibinfo {author} {\bibfnamefont {E.}~\bibnamefont {Grant}}, \bibinfo {author} {\bibfnamefont {L.}~\bibnamefont {Wossnig}}, \bibinfo {author} {\bibfnamefont {I.}~\bibnamefont {Rungger}}, \bibinfo {author} {\bibfnamefont {G.~H.}\ \bibnamefont {Booth}},\ and\ \bibinfo {author} {\bibfnamefont {J.}~\bibnamefont {Tennyson}},\ }\bibfield  {title} {\bibinfo {title} {The variational quantum eigensolver: A review of methods and best practices},\ }\href {https://doi.org/10.1016/j.physrep.2022.08.003} {\bibfield  {journal} {\bibinfo  {journal} {Physics Reports}\ }\textbf {\bibinfo {volume} {986}},\ \bibinfo {pages} {1} (\bibinfo {year}
  {2022})}\BibitemShut {NoStop}%
\bibitem [{\citenamefont {McClean}\ \emph {et~al.}(2018)\citenamefont {McClean}, \citenamefont {Boixo}, \citenamefont {Smelyanskiy}, \citenamefont {Babbush},\ and\ \citenamefont {Neven}}]{McClean2018}%
  \BibitemOpen
  \bibfield  {author} {\bibinfo {author} {\bibfnamefont {J.~R.}\ \bibnamefont {McClean}}, \bibinfo {author} {\bibfnamefont {S.}~\bibnamefont {Boixo}}, \bibinfo {author} {\bibfnamefont {V.~N.}\ \bibnamefont {Smelyanskiy}}, \bibinfo {author} {\bibfnamefont {R.}~\bibnamefont {Babbush}},\ and\ \bibinfo {author} {\bibfnamefont {H.}~\bibnamefont {Neven}},\ }\bibfield  {title} {\bibinfo {title} {Barren plateaus in quantum neural network training landscapes},\ }\href {https://doi.org/10.1038/s41467-018-07090-4} {\bibfield  {journal} {\bibinfo  {journal} {Nature Communications}\ }\textbf {\bibinfo {volume} {9}},\ \bibinfo {pages} {4812} (\bibinfo {year} {2018})}\BibitemShut {NoStop}%
\bibitem [{\citenamefont {Cerezo}\ \emph {et~al.}(2021{\natexlab{b}})\citenamefont {Cerezo}, \citenamefont {Sone}, \citenamefont {Volkoff}, \citenamefont {Cincio},\ and\ \citenamefont {Coles}}]{Cerezo2021Cost}%
  \BibitemOpen
  \bibfield  {author} {\bibinfo {author} {\bibfnamefont {M.}~\bibnamefont {Cerezo}}, \bibinfo {author} {\bibfnamefont {A.}~\bibnamefont {Sone}}, \bibinfo {author} {\bibfnamefont {T.}~\bibnamefont {Volkoff}}, \bibinfo {author} {\bibfnamefont {L.}~\bibnamefont {Cincio}},\ and\ \bibinfo {author} {\bibfnamefont {P.~J.}\ \bibnamefont {Coles}},\ }\bibfield  {title} {\bibinfo {title} {Cost function dependent barren plateaus in shallow parametrized quantum circuits},\ }\href {https://doi.org/10.1038/s41467-021-21728-w} {\bibfield  {journal} {\bibinfo  {journal} {Nature Communications}\ }\textbf {\bibinfo {volume} {12}},\ \bibinfo {pages} {1791} (\bibinfo {year} {2021}{\natexlab{b}})}\BibitemShut {NoStop}%
\bibitem [{\citenamefont {Larocca}\ \emph {et~al.}(2025)\citenamefont {Larocca}, \citenamefont {Thanasilp}, \citenamefont {Wang}, \citenamefont {Sharma}, \citenamefont {Biamonte}, \citenamefont {Coles}, \citenamefont {Cincio}, \citenamefont {McClean}, \citenamefont {Holmes},\ and\ \citenamefont {Cerezo}}]{Larocca2025}%
  \BibitemOpen
  \bibfield  {author} {\bibinfo {author} {\bibfnamefont {M.}~\bibnamefont {Larocca}}, \bibinfo {author} {\bibfnamefont {S.}~\bibnamefont {Thanasilp}}, \bibinfo {author} {\bibfnamefont {S.}~\bibnamefont {Wang}}, \bibinfo {author} {\bibfnamefont {K.}~\bibnamefont {Sharma}}, \bibinfo {author} {\bibfnamefont {J.}~\bibnamefont {Biamonte}}, \bibinfo {author} {\bibfnamefont {P.~J.}\ \bibnamefont {Coles}}, \bibinfo {author} {\bibfnamefont {L.}~\bibnamefont {Cincio}}, \bibinfo {author} {\bibfnamefont {J.~R.}\ \bibnamefont {McClean}}, \bibinfo {author} {\bibfnamefont {Z.}~\bibnamefont {Holmes}},\ and\ \bibinfo {author} {\bibfnamefont {M.}~\bibnamefont {Cerezo}},\ }\bibfield  {title} {\bibinfo {title} {Barren plateaus in variational quantum computing},\ }\href {https://doi.org/10.1038/s42254-025-00813-9} {\bibfield  {journal} {\bibinfo  {journal} {Nature Reviews Physics}\ }\textbf {\bibinfo {volume} {7}},\ \bibinfo {pages} {174} (\bibinfo {year} {2025})}\BibitemShut {NoStop}%
\bibitem [{\citenamefont {Anschuetz}\ and\ \citenamefont {Kiani}(2022)}]{AnschuetzKiani2022}%
  \BibitemOpen
  \bibfield  {author} {\bibinfo {author} {\bibfnamefont {E.~R.}\ \bibnamefont {Anschuetz}}\ and\ \bibinfo {author} {\bibfnamefont {B.~T.}\ \bibnamefont {Kiani}},\ }\bibfield  {title} {\bibinfo {title} {Quantum variational algorithms are swamped with traps},\ }\href {https://doi.org/10.1038/s41467-022-35364-5} {\bibfield  {journal} {\bibinfo  {journal} {Nature Communications}\ }\textbf {\bibinfo {volume} {13}},\ \bibinfo {pages} {7760} (\bibinfo {year} {2022})}\BibitemShut {NoStop}%
\bibitem [{\citenamefont {Nemkov}\ \emph {et~al.}(2025)\citenamefont {Nemkov}, \citenamefont {Kiktenko},\ and\ \citenamefont {Fedorov}}]{Nemkov2025}%
  \BibitemOpen
  \bibfield  {author} {\bibinfo {author} {\bibfnamefont {N.~A.}\ \bibnamefont {Nemkov}}, \bibinfo {author} {\bibfnamefont {E.~O.}\ \bibnamefont {Kiktenko}},\ and\ \bibinfo {author} {\bibfnamefont {A.~K.}\ \bibnamefont {Fedorov}},\ }\bibfield  {title} {\bibinfo {title} {Barren plateaus swamped with traps},\ }\href {https://doi.org/10.1103/PhysRevA.111.012441} {\bibfield  {journal} {\bibinfo  {journal} {Physical Review A}\ }\textbf {\bibinfo {volume} {111}},\ \bibinfo {pages} {012441} (\bibinfo {year} {2025})}\BibitemShut {NoStop}%
\bibitem [{\citenamefont {Bittel}\ and\ \citenamefont {Kliesch}(2021)}]{BittelKliesch2021}%
  \BibitemOpen
  \bibfield  {author} {\bibinfo {author} {\bibfnamefont {L.}~\bibnamefont {Bittel}}\ and\ \bibinfo {author} {\bibfnamefont {M.}~\bibnamefont {Kliesch}},\ }\bibfield  {title} {\bibinfo {title} {Training variational quantum algorithms is np-hard},\ }\href {https://doi.org/10.1103/PhysRevLett.127.120502} {\bibfield  {journal} {\bibinfo  {journal} {Physical Review Letters}\ }\textbf {\bibinfo {volume} {127}},\ \bibinfo {pages} {120502} (\bibinfo {year} {2021})}\BibitemShut {NoStop}%
\bibitem [{\citenamefont {Sack}\ and\ \citenamefont {Serbyn}(2021)}]{SackSerbyn2021}%
  \BibitemOpen
  \bibfield  {author} {\bibinfo {author} {\bibfnamefont {S.~H.}\ \bibnamefont {Sack}}\ and\ \bibinfo {author} {\bibfnamefont {M.}~\bibnamefont {Serbyn}},\ }\bibfield  {title} {\bibinfo {title} {Quantum annealing initialization of the quantum approximate optimization algorithm},\ }\href {https://doi.org/10.22331/q-2021-07-01-491} {\bibfield  {journal} {\bibinfo  {journal} {Quantum}\ }\textbf {\bibinfo {volume} {5}},\ \bibinfo {pages} {491} (\bibinfo {year} {2021})}\BibitemShut {NoStop}%
\bibitem [{\citenamefont {Akshay}\ \emph {et~al.}(2021)\citenamefont {Akshay}, \citenamefont {Philathong}, \citenamefont {Zacharov},\ and\ \citenamefont {Biamonte}}]{Akshay2021}%
  \BibitemOpen
  \bibfield  {author} {\bibinfo {author} {\bibfnamefont {V.}~\bibnamefont {Akshay}}, \bibinfo {author} {\bibfnamefont {H.}~\bibnamefont {Philathong}}, \bibinfo {author} {\bibfnamefont {I.}~\bibnamefont {Zacharov}},\ and\ \bibinfo {author} {\bibfnamefont {J.}~\bibnamefont {Biamonte}},\ }\bibfield  {title} {\bibinfo {title} {Reachability deficits in quantum approximate optimization of graph problems},\ }\href {https://doi.org/10.22331/q-2021-08-30-532} {\bibfield  {journal} {\bibinfo  {journal} {Quantum}\ }\textbf {\bibinfo {volume} {5}},\ \bibinfo {pages} {532} (\bibinfo {year} {2021})}\BibitemShut {NoStop}%
\bibitem [{\citenamefont {Boy}\ and\ \citenamefont {Wales}(2024)}]{BoyWales2024}%
  \BibitemOpen
  \bibfield  {author} {\bibinfo {author} {\bibfnamefont {C.}~\bibnamefont {Boy}}\ and\ \bibinfo {author} {\bibfnamefont {D.~J.}\ \bibnamefont {Wales}},\ }\bibfield  {title} {\bibinfo {title} {Energy landscapes for the quantum approximate optimization algorithm},\ }\href {https://doi.org/10.1103/PhysRevA.109.062602} {\bibfield  {journal} {\bibinfo  {journal} {Physical Review A}\ }\textbf {\bibinfo {volume} {109}},\ \bibinfo {pages} {062602} (\bibinfo {year} {2024})}\BibitemShut {NoStop}%
\bibitem [{\citenamefont {Storn}\ and\ \citenamefont {Price}(1997)}]{StornPrice1997}%
  \BibitemOpen
  \bibfield  {author} {\bibinfo {author} {\bibfnamefont {R.}~\bibnamefont {Storn}}\ and\ \bibinfo {author} {\bibfnamefont {K.}~\bibnamefont {Price}},\ }\bibfield  {title} {\bibinfo {title} {Differential evolution---a simple and efficient heuristic for global optimization over continuous spaces},\ }\href {https://doi.org/10.1023/A:1008202821328} {\bibfield  {journal} {\bibinfo  {journal} {Journal of Global Optimization}\ }\textbf {\bibinfo {volume} {11}},\ \bibinfo {pages} {341} (\bibinfo {year} {1997})}\BibitemShut {NoStop}%
\bibitem [{\citenamefont {Fa{\'i}lde}\ \emph {et~al.}(2023)\citenamefont {Fa{\'i}lde}, \citenamefont {Viqueira}, \citenamefont {Mussa~Juane},\ and\ \citenamefont {G{\'o}mez}}]{Failde2023}%
  \BibitemOpen
  \bibfield  {author} {\bibinfo {author} {\bibfnamefont {D.}~\bibnamefont {Fa{\'i}lde}}, \bibinfo {author} {\bibfnamefont {J.~D.}\ \bibnamefont {Viqueira}}, \bibinfo {author} {\bibfnamefont {M.}~\bibnamefont {Mussa~Juane}},\ and\ \bibinfo {author} {\bibfnamefont {A.}~\bibnamefont {G{\'o}mez}},\ }\bibfield  {title} {\bibinfo {title} {Using differential evolution to avoid local minima in variational quantum algorithms},\ }\href {https://doi.org/10.1038/s41598-023-43404-3} {\bibfield  {journal} {\bibinfo  {journal} {Scientific Reports}\ }\textbf {\bibinfo {volume} {13}},\ \bibinfo {pages} {16230} (\bibinfo {year} {2023})}\BibitemShut {NoStop}%
\bibitem [{\citenamefont {Mei}\ \emph {et~al.}(2024)\citenamefont {Mei}, \citenamefont {Zhao}, \citenamefont {Li}, \citenamefont {Chen}, \citenamefont {Zhang}, \citenamefont {Wang}, \citenamefont {Wu},\ and\ \citenamefont {Guo}}]{Mei2024}%
  \BibitemOpen
  \bibfield  {author} {\bibinfo {author} {\bibfnamefont {H.}~\bibnamefont {Mei}}, \bibinfo {author} {\bibfnamefont {J.}~\bibnamefont {Zhao}}, \bibinfo {author} {\bibfnamefont {Q.-S.}\ \bibnamefont {Li}}, \bibinfo {author} {\bibfnamefont {Z.-Y.}\ \bibnamefont {Chen}}, \bibinfo {author} {\bibfnamefont {J.-J.}\ \bibnamefont {Zhang}}, \bibinfo {author} {\bibfnamefont {Q.}~\bibnamefont {Wang}}, \bibinfo {author} {\bibfnamefont {Y.-C.}\ \bibnamefont {Wu}},\ and\ \bibinfo {author} {\bibfnamefont {G.-P.}\ \bibnamefont {Guo}},\ }\bibfield  {title} {\bibinfo {title} {Particle swarm optimization for a variational quantum eigensolver},\ }\href {https://doi.org/10.1039/D4CP02021A} {\bibfield  {journal} {\bibinfo  {journal} {Physical Chemistry Chemical Physics}\ }\textbf {\bibinfo {volume} {26}},\ \bibinfo {pages} {29070} (\bibinfo {year} {2024})}\BibitemShut {NoStop}%
\bibitem [{\citenamefont {Jones}\ \emph {et~al.}(2025)\citenamefont {Jones}, \citenamefont {Mineh},\ and\ \citenamefont {Montanaro}}]{Jones2025}%
  \BibitemOpen
  \bibfield  {author} {\bibinfo {author} {\bibfnamefont {B.~D.~M.}\ \bibnamefont {Jones}}, \bibinfo {author} {\bibfnamefont {L.}~\bibnamefont {Mineh}},\ and\ \bibinfo {author} {\bibfnamefont {A.}~\bibnamefont {Montanaro}},\ }\bibfield  {title} {\bibinfo {title} {Benchmarking a wide range of optimisers for solving the fermi--hubbard model using the variational quantum eigensolver},\ }\href {https://doi.org/10.1088/2058-9565/adfe15} {\bibfield  {journal} {\bibinfo  {journal} {Quantum Science and Technology}\ }\textbf {\bibinfo {volume} {10}},\ \bibinfo {pages} {045032} (\bibinfo {year} {2025})},\ \Eprint {https://arxiv.org/abs/2411.13742} {arXiv:2411.13742} \BibitemShut {NoStop}%
\bibitem [{\citenamefont {Ill{\'e}sov{\'a}}\ \emph {et~al.}(2025{\natexlab{a}})\citenamefont {Ill{\'e}sov{\'a}}, \citenamefont {Rybotycki},\ and\ \citenamefont {Beseda}}]{illesova2025qmetric}%
  \BibitemOpen
  \bibfield  {author} {\bibinfo {author} {\bibfnamefont {S.}~\bibnamefont {Ill{\'e}sov{\'a}}}, \bibinfo {author} {\bibfnamefont {T.}~\bibnamefont {Rybotycki}},\ and\ \bibinfo {author} {\bibfnamefont {M.}~\bibnamefont {Beseda}},\ }\bibfield  {title} {\bibinfo {title} {Qmetric: Benchmarking quantum neural networks across circuits, features, and training dimensions},\ }in\ \href {https://ceur-ws.org/Vol-4080/paper3.pdf} {\emph {\bibinfo {booktitle} {Proceedings of QualITA 2025: The Fourth Conference on System and Service Quality}}},\ \bibinfo {series} {CEUR Workshop Proceedings}, Vol.\ \bibinfo {volume} {4080}\ (\bibinfo  {publisher} {CEUR-WS.org},\ \bibinfo {address} {Catania, Italy},\ \bibinfo {year} {2025})\BibitemShut {NoStop}%
\bibitem [{\citenamefont {Ill{\'e}sov{\'a}}\ \emph {et~al.}(2025{\natexlab{b}})\citenamefont {Ill{\'e}sov{\'a}}, \citenamefont {Bezd{\v{e}}k}, \citenamefont {Nov{\'a}k}, \citenamefont {Senjean},\ and\ \citenamefont {Beseda}}]{Illesova2025Statistical}%
  \BibitemOpen
  \bibfield  {author} {\bibinfo {author} {\bibfnamefont {S.}~\bibnamefont {Ill{\'e}sov{\'a}}}, \bibinfo {author} {\bibfnamefont {T.}~\bibnamefont {Bezd{\v{e}}k}}, \bibinfo {author} {\bibfnamefont {V.}~\bibnamefont {Nov{\'a}k}}, \bibinfo {author} {\bibfnamefont {B.}~\bibnamefont {Senjean}},\ and\ \bibinfo {author} {\bibfnamefont {M.}~\bibnamefont {Beseda}},\ }\href@noop {} {\bibinfo {title} {Statistical benchmarking of optimization methods for variational quantum eigensolver under quantum noise}} (\bibinfo {year} {2025}{\natexlab{b}}),\ \Eprint {https://arxiv.org/abs/2510.08727} {arXiv:2510.08727 [quant-ph]} \BibitemShut {NoStop}%
\bibitem [{\citenamefont {Bonet-Monroig}\ \emph {et~al.}(2023)\citenamefont {Bonet-Monroig}, \citenamefont {Wang}, \citenamefont {Vermetten}, \citenamefont {Senjean}, \citenamefont {Moussa}, \citenamefont {B{\"a}ck}, \citenamefont {Dunjko},\ and\ \citenamefont {O'Brien}}]{bonet2023performance}%
  \BibitemOpen
  \bibfield  {author} {\bibinfo {author} {\bibfnamefont {X.}~\bibnamefont {Bonet-Monroig}}, \bibinfo {author} {\bibfnamefont {H.}~\bibnamefont {Wang}}, \bibinfo {author} {\bibfnamefont {D.}~\bibnamefont {Vermetten}}, \bibinfo {author} {\bibfnamefont {B.}~\bibnamefont {Senjean}}, \bibinfo {author} {\bibfnamefont {C.}~\bibnamefont {Moussa}}, \bibinfo {author} {\bibfnamefont {T.}~\bibnamefont {B{\"a}ck}}, \bibinfo {author} {\bibfnamefont {V.}~\bibnamefont {Dunjko}},\ and\ \bibinfo {author} {\bibfnamefont {T.~E.}\ \bibnamefont {O'Brien}},\ }\bibfield  {title} {\bibinfo {title} {Performance comparison of optimization methods on variational quantum algorithms},\ }\href@noop {} {\bibfield  {journal} {\bibinfo  {journal} {Physical Review A}\ }\textbf {\bibinfo {volume} {107}},\ \bibinfo {pages} {032407} (\bibinfo {year} {2023})}\BibitemShut {NoStop}%
\bibitem [{\citenamefont {Bezd{\v{e}}k}\ \emph {et~al.}(2025)\citenamefont {Bezd{\v{e}}k}, \citenamefont {Yuan}, \citenamefont {Nov{\'a}k}, \citenamefont {Ill{\'e}sov{\'a}},\ and\ \citenamefont {Beseda}}]{Bezdek2025ClassicalOptimization}%
  \BibitemOpen
  \bibfield  {author} {\bibinfo {author} {\bibfnamefont {T.}~\bibnamefont {Bezd{\v{e}}k}}, \bibinfo {author} {\bibfnamefont {H.}~\bibnamefont {Yuan}}, \bibinfo {author} {\bibfnamefont {V.}~\bibnamefont {Nov{\'a}k}}, \bibinfo {author} {\bibfnamefont {S.}~\bibnamefont {Ill{\'e}sov{\'a}}},\ and\ \bibinfo {author} {\bibfnamefont {M.}~\bibnamefont {Beseda}},\ }\href@noop {} {\bibinfo {title} {Classical optimization strategies for variational quantum algorithms: A systematic study of noise effects and parameter efficiency}} (\bibinfo {year} {2025}),\ \Eprint {https://arxiv.org/abs/2511.09314} {arXiv:2511.09314 [quant-ph]} \BibitemShut {NoStop}%
\bibitem [{\citenamefont {Nov{\'a}k}\ \emph {et~al.}(2025)\citenamefont {Nov{\'a}k}, \citenamefont {Ill{\'e}sov{\'a}}, \citenamefont {Bezd{\v{e}}k}, \citenamefont {Zelinka},\ and\ \citenamefont {Beseda}}]{Novak2025Reliable}%
  \BibitemOpen
  \bibfield  {author} {\bibinfo {author} {\bibfnamefont {V.}~\bibnamefont {Nov{\'a}k}}, \bibinfo {author} {\bibfnamefont {S.}~\bibnamefont {Ill{\'e}sov{\'a}}}, \bibinfo {author} {\bibfnamefont {T.}~\bibnamefont {Bezd{\v{e}}k}}, \bibinfo {author} {\bibfnamefont {I.}~\bibnamefont {Zelinka}},\ and\ \bibinfo {author} {\bibfnamefont {M.}~\bibnamefont {Beseda}},\ }\href@noop {} {\bibinfo {title} {Reliable optimization under noise in quantum variational algorithms}} (\bibinfo {year} {2025}),\ \Eprint {https://arxiv.org/abs/2511.08289} {arXiv:2511.08289 [quant-ph]} \BibitemShut {NoStop}%
\bibitem [{\citenamefont {Nov{\'a}k}\ \emph {et~al.}(2026{\natexlab{a}})\citenamefont {Nov{\'a}k}, \citenamefont {Zelinka},\ and\ \citenamefont {Sn{\'a}{\v{s}}el}}]{Novak2025NoisyLandscapes}%
  \BibitemOpen
  \bibfield  {author} {\bibinfo {author} {\bibfnamefont {V.}~\bibnamefont {Nov{\'a}k}}, \bibinfo {author} {\bibfnamefont {I.}~\bibnamefont {Zelinka}},\ and\ \bibinfo {author} {\bibfnamefont {V.}~\bibnamefont {Sn{\'a}{\v{s}}el}},\ }\bibfield  {title} {\bibinfo {title} {Optimization strategies for variational quantum algorithms in noisy landscapes},\ }\href {https://doi.org/10.1007/s12065-026-01248-6} {\bibfield  {journal} {\bibinfo  {journal} {Evolutionary Intelligence}\ }\textbf {\bibinfo {volume} {19}},\ \bibinfo {pages} {142} (\bibinfo {year} {2026}{\natexlab{a}})}\BibitemShut {NoStop}%
\bibitem [{\citenamefont {Hansen}\ \emph {et~al.}(2010)\citenamefont {Hansen}, \citenamefont {Auger}, \citenamefont {Ros}, \citenamefont {Finck},\ and\ \citenamefont {Po{\v{s}}{\'\i}k}}]{hansen2010comparing}%
  \BibitemOpen
  \bibfield  {author} {\bibinfo {author} {\bibfnamefont {N.}~\bibnamefont {Hansen}}, \bibinfo {author} {\bibfnamefont {A.}~\bibnamefont {Auger}}, \bibinfo {author} {\bibfnamefont {R.}~\bibnamefont {Ros}}, \bibinfo {author} {\bibfnamefont {S.}~\bibnamefont {Finck}},\ and\ \bibinfo {author} {\bibfnamefont {P.}~\bibnamefont {Po{\v{s}}{\'\i}k}},\ }\bibfield  {title} {\bibinfo {title} {Comparing results of 31 algorithms from the black-box optimization benchmarking bbob-2009},\ }in\ \href@noop {} {\emph {\bibinfo {booktitle} {Proceedings of the 12th annual conference companion on Genetic and evolutionary computation}}}\ (\bibinfo {year} {2010})\ pp.\ \bibinfo {pages} {1689--1696}\BibitemShut {NoStop}%
\bibitem [{\citenamefont {Nov{\'a}k}\ \emph {et~al.}(2026{\natexlab{b}})\citenamefont {Nov{\'a}k}, \citenamefont {Bezd{\v{e}}k}, \citenamefont {Zelinka}, \citenamefont {Das},\ and\ \citenamefont {Beseda}}]{Novak2026CEC}%
  \BibitemOpen
  \bibfield  {author} {\bibinfo {author} {\bibfnamefont {V.}~\bibnamefont {Nov{\'a}k}}, \bibinfo {author} {\bibfnamefont {T.}~\bibnamefont {Bezd{\v{e}}k}}, \bibinfo {author} {\bibfnamefont {I.}~\bibnamefont {Zelinka}}, \bibinfo {author} {\bibfnamefont {S.}~\bibnamefont {Das}},\ and\ \bibinfo {author} {\bibfnamefont {M.}~\bibnamefont {Beseda}},\ }\bibfield  {title} {\bibinfo {title} {A longitudinal analysis of the cec single-objective competitions (2010--2024) and implications for variational quantum optimization},\ }\href {https://doi.org/10.1016/j.swevo.2026.102469} {\bibfield  {journal} {\bibinfo  {journal} {Swarm and Evolutionary Computation}\ }\textbf {\bibinfo {volume} {107}},\ \bibinfo {pages} {102469} (\bibinfo {year} {2026}{\natexlab{b}})}\BibitemShut {NoStop}%
\bibitem [{\citenamefont {Finck}\ \emph {et~al.}(2010)\citenamefont {Finck}, \citenamefont {Hansen}, \citenamefont {Ros},\ and\ \citenamefont {Auger}}]{finck2010}%
  \BibitemOpen
  \bibfield  {author} {\bibinfo {author} {\bibfnamefont {S.}~\bibnamefont {Finck}}, \bibinfo {author} {\bibfnamefont {N.}~\bibnamefont {Hansen}}, \bibinfo {author} {\bibfnamefont {R.}~\bibnamefont {Ros}},\ and\ \bibinfo {author} {\bibfnamefont {A.}~\bibnamefont {Auger}},\ }\href@noop {} {\emph {\bibinfo {title} {Real-Parameter Black-Box Optimization Benchmarking 2010: Presentation of the Noiseless Functions}}},\ \bibinfo {type} {Tech. Rep.}\ \bibinfo {number} {2009/20}\ (\bibinfo  {institution} {Research Center PPE, University of Applied Sciences Vorarlberg},\ \bibinfo {year} {2010})\BibitemShut {NoStop}%
\bibitem [{\citenamefont {Awad}\ \emph {et~al.}(2016)\citenamefont {Awad}, \citenamefont {Ali}, \citenamefont {Suganthan}, \citenamefont {Liang},\ and\ \citenamefont {Qu}}]{awad2017}%
  \BibitemOpen
  \bibfield  {author} {\bibinfo {author} {\bibfnamefont {N.~H.}\ \bibnamefont {Awad}}, \bibinfo {author} {\bibfnamefont {M.~Z.}\ \bibnamefont {Ali}}, \bibinfo {author} {\bibfnamefont {P.~N.}\ \bibnamefont {Suganthan}}, \bibinfo {author} {\bibfnamefont {J.~J.}\ \bibnamefont {Liang}},\ and\ \bibinfo {author} {\bibfnamefont {B.~Y.}\ \bibnamefont {Qu}},\ }\href@noop {} {\emph {\bibinfo {title} {Problem Definitions and Evaluation Criteria for the CEC 2017 Special Session and Competition on Single Objective Real-Parameter Numerical Optimization}}},\ \bibinfo {type} {Tech. Rep.}\ (\bibinfo  {institution} {Nanyang Technological University, Singapore},\ \bibinfo {year} {2016})\BibitemShut {NoStop}%
\bibitem [{\citenamefont {Piotrowski}\ \emph {et~al.}(2023)\citenamefont {Piotrowski}, \citenamefont {Napiorkowski},\ and\ \citenamefont {Piotrowska}}]{Piotrowski2023}%
  \BibitemOpen
  \bibfield  {author} {\bibinfo {author} {\bibfnamefont {A.~P.}\ \bibnamefont {Piotrowski}}, \bibinfo {author} {\bibfnamefont {J.~J.}\ \bibnamefont {Napiorkowski}},\ and\ \bibinfo {author} {\bibfnamefont {A.~E.}\ \bibnamefont {Piotrowska}},\ }\bibfield  {title} {\bibinfo {title} {Choice of benchmark optimization problems does matter},\ }\href {https://doi.org/10.1016/j.swevo.2023.101378} {\bibfield  {journal} {\bibinfo  {journal} {Swarm and Evolutionary Computation}\ }\textbf {\bibinfo {volume} {83}},\ \bibinfo {pages} {101378} (\bibinfo {year} {2023})}\BibitemShut {NoStop}%
\bibitem [{\citenamefont {LaTorre}\ \emph {et~al.}(2021)\citenamefont {LaTorre}, \citenamefont {Molina}, \citenamefont {Osaba}, \citenamefont {Poyatos}, \citenamefont {Del~Ser},\ and\ \citenamefont {Herrera}}]{LaTorre2021}%
  \BibitemOpen
  \bibfield  {author} {\bibinfo {author} {\bibfnamefont {A.}~\bibnamefont {LaTorre}}, \bibinfo {author} {\bibfnamefont {D.}~\bibnamefont {Molina}}, \bibinfo {author} {\bibfnamefont {E.}~\bibnamefont {Osaba}}, \bibinfo {author} {\bibfnamefont {J.}~\bibnamefont {Poyatos}}, \bibinfo {author} {\bibfnamefont {J.}~\bibnamefont {Del~Ser}},\ and\ \bibinfo {author} {\bibfnamefont {F.}~\bibnamefont {Herrera}},\ }\bibfield  {title} {\bibinfo {title} {A prescription of methodological guidelines for comparing bio-inspired optimization algorithms},\ }\href {https://doi.org/10.1016/j.swevo.2021.100973} {\bibfield  {journal} {\bibinfo  {journal} {Swarm and Evolutionary Computation}\ }\textbf {\bibinfo {volume} {67}},\ \bibinfo {pages} {100973} (\bibinfo {year} {2021})}\BibitemShut {NoStop}%
\bibitem [{\citenamefont {Nikolikj}\ \emph {et~al.}(2025)\citenamefont {Nikolikj}, \citenamefont {Mu{\~n}oz},\ and\ \citenamefont {Eftimov}}]{Nikolikj2025}%
  \BibitemOpen
  \bibfield  {author} {\bibinfo {author} {\bibfnamefont {A.}~\bibnamefont {Nikolikj}}, \bibinfo {author} {\bibfnamefont {M.~A.}\ \bibnamefont {Mu{\~n}oz}},\ and\ \bibinfo {author} {\bibfnamefont {T.}~\bibnamefont {Eftimov}},\ }\bibfield  {title} {\bibinfo {title} {Benchmarking footprints of continuous black-box optimization algorithms: Explainable insights into algorithm success and failure},\ }\href {https://doi.org/10.1016/j.swevo.2025.101895} {\bibfield  {journal} {\bibinfo  {journal} {Swarm and Evolutionary Computation}\ }\textbf {\bibinfo {volume} {94}},\ \bibinfo {pages} {101895} (\bibinfo {year} {2025})}\BibitemShut {NoStop}%
\bibitem [{\citenamefont {Cenikj}\ \emph {et~al.}(2026)\citenamefont {Cenikj}, \citenamefont {Nikolikj}, \citenamefont {Petelin}, \citenamefont {van Stein}, \citenamefont {Doerr},\ and\ \citenamefont {Eftimov}}]{Cenikj2026}%
  \BibitemOpen
  \bibfield  {author} {\bibinfo {author} {\bibfnamefont {G.}~\bibnamefont {Cenikj}}, \bibinfo {author} {\bibfnamefont {A.}~\bibnamefont {Nikolikj}}, \bibinfo {author} {\bibfnamefont {G.}~\bibnamefont {Petelin}}, \bibinfo {author} {\bibfnamefont {N.}~\bibnamefont {van Stein}}, \bibinfo {author} {\bibfnamefont {C.}~\bibnamefont {Doerr}},\ and\ \bibinfo {author} {\bibfnamefont {T.}~\bibnamefont {Eftimov}},\ }\bibfield  {title} {\bibinfo {title} {A survey of features used for representing black-box single-objective continuous optimization},\ }\href {https://doi.org/10.1016/j.swevo.2026.102288} {\bibfield  {journal} {\bibinfo  {journal} {Swarm and Evolutionary Computation}\ }\textbf {\bibinfo {volume} {101}},\ \bibinfo {pages} {102288} (\bibinfo {year} {2026})}\BibitemShut {NoStop}%
\bibitem [{\citenamefont {Nov{\'a}k}\ \emph {et~al.}(2026{\natexlab{c}})\citenamefont {Nov{\'a}k}, \citenamefont {Zelinka}, \citenamefont {Das},\ and\ \citenamefont {Beseda}}]{Novak2026Landscape}%
  \BibitemOpen
  \bibfield  {author} {\bibinfo {author} {\bibfnamefont {V.}~\bibnamefont {Nov{\'a}k}}, \bibinfo {author} {\bibfnamefont {I.}~\bibnamefont {Zelinka}}, \bibinfo {author} {\bibfnamefont {S.}~\bibnamefont {Das}},\ and\ \bibinfo {author} {\bibfnamefont {M.}~\bibnamefont {Beseda}},\ }\href@noop {} {\bibinfo {title} {Optimization landscape geometry in {VQE} for frustrated quantum spin models}} (\bibinfo {year} {2026}{\natexlab{c}}),\ \Eprint {https://arxiv.org/abs/2609.00235} {arXiv:2609.00235 [quant-ph]} \BibitemShut {NoStop}%
\bibitem [{\citenamefont {Cenikj}\ \emph {et~al.}(2025)\citenamefont {Cenikj}, \citenamefont {Petelin}, \citenamefont {Seiler}, \citenamefont {Cenikj},\ and\ \citenamefont {Eftimov}}]{Cenikj2025}%
  \BibitemOpen
  \bibfield  {author} {\bibinfo {author} {\bibfnamefont {G.}~\bibnamefont {Cenikj}}, \bibinfo {author} {\bibfnamefont {G.}~\bibnamefont {Petelin}}, \bibinfo {author} {\bibfnamefont {M.}~\bibnamefont {Seiler}}, \bibinfo {author} {\bibfnamefont {N.}~\bibnamefont {Cenikj}},\ and\ \bibinfo {author} {\bibfnamefont {T.}~\bibnamefont {Eftimov}},\ }\bibfield  {title} {\bibinfo {title} {Landscape features in single-objective continuous optimization: Have we hit a wall in algorithm selection generalization?},\ }\href {https://doi.org/10.1016/j.swevo.2025.101894} {\bibfield  {journal} {\bibinfo  {journal} {Swarm and Evolutionary Computation}\ }\textbf {\bibinfo {volume} {94}},\ \bibinfo {pages} {101894} (\bibinfo {year} {2025})}\BibitemShut {NoStop}%
\bibitem [{\citenamefont {Nocedal}\ and\ \citenamefont {Wright}(2006)}]{NocedalWright2006}%
  \BibitemOpen
  \bibfield  {author} {\bibinfo {author} {\bibfnamefont {J.}~\bibnamefont {Nocedal}}\ and\ \bibinfo {author} {\bibfnamefont {S.~J.}\ \bibnamefont {Wright}},\ }\href@noop {} {\emph {\bibinfo {title} {Numerical Optimization}}},\ \bibinfo {edition} {2nd}\ ed.\ (\bibinfo  {publisher} {Springer},\ \bibinfo {year} {2006})\BibitemShut {NoStop}%
\bibitem [{\citenamefont {Hansen}\ and\ \citenamefont {Ostermeier}(2001)}]{HansenOstermeier2001}%
  \BibitemOpen
  \bibfield  {author} {\bibinfo {author} {\bibfnamefont {N.}~\bibnamefont {Hansen}}\ and\ \bibinfo {author} {\bibfnamefont {A.}~\bibnamefont {Ostermeier}},\ }\bibfield  {title} {\bibinfo {title} {Completely derandomized self-adaptation in evolution strategies},\ }\href {https://doi.org/10.1162/106365601750190398} {\bibfield  {journal} {\bibinfo  {journal} {Evolutionary Computation}\ }\textbf {\bibinfo {volume} {9}},\ \bibinfo {pages} {159} (\bibinfo {year} {2001})}\BibitemShut {NoStop}%
\bibitem [{\citenamefont {Tanabe}\ and\ \citenamefont {Fukunaga}(2013)}]{TanabeFukunaga2013}%
  \BibitemOpen
  \bibfield  {author} {\bibinfo {author} {\bibfnamefont {R.}~\bibnamefont {Tanabe}}\ and\ \bibinfo {author} {\bibfnamefont {A.}~\bibnamefont {Fukunaga}},\ }\bibfield  {title} {\bibinfo {title} {Success-history based parameter adaptation for differential evolution},\ }in\ \href {https://doi.org/10.1109/CEC.2013.6557555} {\emph {\bibinfo {booktitle} {2013 IEEE Congress on Evolutionary Computation}}}\ (\bibinfo {organization} {IEEE},\ \bibinfo {year} {2013})\ pp.\ \bibinfo {pages} {71--78}\BibitemShut {NoStop}%
\bibitem [{\citenamefont {Tanabe}\ and\ \citenamefont {Fukunaga}(2014)}]{TanabeFukunaga2014}%
  \BibitemOpen
  \bibfield  {author} {\bibinfo {author} {\bibfnamefont {R.}~\bibnamefont {Tanabe}}\ and\ \bibinfo {author} {\bibfnamefont {A.~S.}\ \bibnamefont {Fukunaga}},\ }\bibfield  {title} {\bibinfo {title} {Improving the search performance of shade using linear population size reduction},\ }in\ \href {https://doi.org/10.1109/CEC.2014.6900380} {\emph {\bibinfo {booktitle} {2014 IEEE Congress on Evolutionary Computation}}}\ (\bibinfo {organization} {IEEE},\ \bibinfo {year} {2014})\ pp.\ \bibinfo {pages} {1658--1665}\BibitemShut {NoStop}%
\bibitem [{\citenamefont {Brest}\ \emph {et~al.}(2016)\citenamefont {Brest}, \citenamefont {Sepesy~Mau{\v{c}}ec},\ and\ \citenamefont {Bo{\v{s}}kovi{\'c}}}]{Brest2016iLSHADE}%
  \BibitemOpen
  \bibfield  {author} {\bibinfo {author} {\bibfnamefont {J.}~\bibnamefont {Brest}}, \bibinfo {author} {\bibfnamefont {M.}~\bibnamefont {Sepesy~Mau{\v{c}}ec}},\ and\ \bibinfo {author} {\bibfnamefont {B.}~\bibnamefont {Bo{\v{s}}kovi{\'c}}},\ }\bibfield  {title} {\bibinfo {title} {il-shade: Improved l-shade algorithm for single objective real-parameter optimization},\ }in\ \href {https://doi.org/10.1109/CEC.2016.7743922} {\emph {\bibinfo {booktitle} {2016 IEEE Congress on Evolutionary Computation (CEC)}}}\ (\bibinfo {organization} {IEEE},\ \bibinfo {year} {2016})\ pp.\ \bibinfo {pages} {1188--1195}\BibitemShut {NoStop}%
\bibitem [{\citenamefont {Brest}\ \emph {et~al.}(2017)\citenamefont {Brest}, \citenamefont {Sepesy~Mau{\v{c}}ec},\ and\ \citenamefont {Bo{\v{s}}kovi{\'c}}}]{Brest2017jSO}%
  \BibitemOpen
  \bibfield  {author} {\bibinfo {author} {\bibfnamefont {J.}~\bibnamefont {Brest}}, \bibinfo {author} {\bibfnamefont {M.}~\bibnamefont {Sepesy~Mau{\v{c}}ec}},\ and\ \bibinfo {author} {\bibfnamefont {B.}~\bibnamefont {Bo{\v{s}}kovi{\'c}}},\ }\bibfield  {title} {\bibinfo {title} {Single objective real-parameter optimization: Algorithm jso},\ }in\ \href {https://doi.org/10.1109/CEC.2017.7969456} {\emph {\bibinfo {booktitle} {2017 IEEE Congress on Evolutionary Computation (CEC)}}}\ (\bibinfo {organization} {IEEE},\ \bibinfo {year} {2017})\ pp.\ \bibinfo {pages} {1311--1318}\BibitemShut {NoStop}%
\bibitem [{\citenamefont {Stanovov}\ and\ \citenamefont {Semenkin}(2024)}]{Stanovov2024LSRTDE}%
  \BibitemOpen
  \bibfield  {author} {\bibinfo {author} {\bibfnamefont {V.}~\bibnamefont {Stanovov}}\ and\ \bibinfo {author} {\bibfnamefont {E.}~\bibnamefont {Semenkin}},\ }\bibfield  {title} {\bibinfo {title} {Success rate-based adaptive differential evolution l-srtde for cec 2024 competition},\ }in\ \href {https://doi.org/10.1109/CEC60901.2024.10611907} {\emph {\bibinfo {booktitle} {2024 IEEE Congress on Evolutionary Computation (CEC)}}}\ (\bibinfo {organization} {IEEE},\ \bibinfo {year} {2024})\ pp.\ \bibinfo {pages} {1--8}\BibitemShut {NoStop}%
\bibitem [{\citenamefont {Schuld}\ \emph {et~al.}(2021)\citenamefont {Schuld}, \citenamefont {Sweke},\ and\ \citenamefont {Meyer}}]{Schuld2021}%
  \BibitemOpen
  \bibfield  {author} {\bibinfo {author} {\bibfnamefont {M.}~\bibnamefont {Schuld}}, \bibinfo {author} {\bibfnamefont {R.}~\bibnamefont {Sweke}},\ and\ \bibinfo {author} {\bibfnamefont {J.~J.}\ \bibnamefont {Meyer}},\ }\bibfield  {title} {\bibinfo {title} {Effect of data encoding on the expressive power of variational quantum-machine-learning models},\ }\href {https://doi.org/10.1103/PhysRevA.103.032430} {\bibfield  {journal} {\bibinfo  {journal} {Physical Review A}\ }\textbf {\bibinfo {volume} {103}},\ \bibinfo {pages} {032430} (\bibinfo {year} {2021})}\BibitemShut {NoStop}%
\bibitem [{\citenamefont {Powell}(1964)}]{Powell1964}%
  \BibitemOpen
  \bibfield  {author} {\bibinfo {author} {\bibfnamefont {M.~J.~D.}\ \bibnamefont {Powell}},\ }\bibfield  {title} {\bibinfo {title} {An efficient method for finding the minimum of a function of several variables without calculating derivatives},\ }\href {https://doi.org/10.1093/comjnl/7.2.155} {\bibfield  {journal} {\bibinfo  {journal} {The Computer Journal}\ }\textbf {\bibinfo {volume} {7}},\ \bibinfo {pages} {155} (\bibinfo {year} {1964})}\BibitemShut {NoStop}%
\end{thebibliography}%

\end{document}